\documentclass[aps,pra,floatfix,superscriptaddress,showpacs]{revtex4} %preprint

\usepackage{graphicx, amsmath, amssymb, booktabs}

\newcommand{\al}{\alpha}     
\newcommand{\be}{\beta}      
\newcommand{\ga}{\gamma}         
\newcommand{\de}{\delta}   \newcommand{\De}{\Delta}      
\newcommand{\ep}{\epsilon}   

\newcommand{\ze}{\zeta}      
       
\newcommand{\thet}{\theta}

\newcommand{\ka}{\kappa}     
\newcommand{\la}{\lambda}

\newcommand{\rh}{\rho}

\newcommand{\ta}{\tau}       
     
\newcommand{\ph}{\phi}     \newcommand{\Ph}{\Phi}        \newcommand{\iPh}{\varPhi}

\newcommand{\ch}{\chi}       
\newcommand{\ps}{\psi}             
\newcommand{\om}{\omega}         \newcommand{\iOm}{\varOmega}

\newcommand{\imp}{\Rightarrow}

\newcommand{\eq}[2][]{\begin{equation}\begin{aligned}#2\end{aligned}\label{#1}\end{equation}}

\newcommand{\eqa}[2][]{\eq{\begin{alignedat}{9}#2\end{alignedat}\label{#1}}}

\newcommand{\fig}[5][1]{\begin{figure}[!htbp]\begin{center}\includegraphics[scale=#1]{#3}\caption[#5]{#4}\label{#2}\end{center}\end{figure}}

\newcommand{\nm}[1]{\!\begin{array}{*{20}c}#1\end{array}\!\!} % ### IOP fix
\newcommand{\bm}[1]{\left[\begin{array}{*{20}c}#1\end{array}\right]} % ### IOP fix

\newcommand{\p}[1]{\left(#1\right)}
\newcommand{\pb}[1]{\left[#1\right]}
\newcommand{\pB}[1]{\left\{#1\right\}}
\newcommand{\pv}[1]{\left|#1\right|}

\newcommand{\<}[1]{\p{#1}}

\newcommand{\C}[2]{\pb{#1,#2}}
\newcommand{\PB}[3][]{\pB{#2,#3}_{#1}}

\newcommand{\E}[1]{\left<#1\right>}

\newcommand{\qo}[2][]{{\hat{#2}{}}_{#1}} % possibly need to remove ...
\newcommand{\ma}[1]{\mathbf{#1}}
\newcommand{\gma}[1]{\boldsymbol{#1}}

\newcommand{\un}[1]{\mathrm{#1}}

\newcommand{\txt}[1]{\mathrm{#1}} % ### IOP fix

\newcommand{\ham}{\mathcal{H}}
\newcommand{\qham}[1][]{\qo[#1]{\ham}}

\newcommand{\ee}[1][]{{\mathrm{e}^{#1}}}
\newcommand{\ii}{\mathrm{i}}

\newcommand{\hb}{\hbar}

\newcommand{\mt}{^{\txt{T}}}
\newcommand{\ct}{^{\dagger}}

\newcommand{\pr}{^{\prime}}

\newcommand{\abs}[1]{\pv{#1}}

\newcommand{\dd}{\mathrm{d}}

\newcommand{\pd}{\partial}
\newcommand{\D}[3][]{\frac{\dd^{#1}#2}{\dd {#3}^{#1}}}
\newcommand{\Dt}[2][]{\D[#1]{#2}{t}}

\newcommand{\pD}[3][]{\frac{\pd^{#1}#2}{\pd {#3}^{#1}}}

\newcommand{\dm}{{\qo{\rh}}}
\newcommand{\Iop}[1][]{{\qo[#1]{I}}}
\newcommand{\qop}[1][]{{\qo[#1]{q}}}
\newcommand{\pop}[1][]{{\qo[#1]{p}}}
\newcommand{\xop}[1][]{{\qo[#1]{x}}}
\newcommand{\yop}[1][]{{\qo[#1]{y}}}
\newcommand{\aop}[1][]{{\qo[#1]{a}}}
\newcommand{\bop}[1][]{{\qo[#1]{b}}}

\newcommand{\Bop}[1][]{{\qo[#1]{B}}}
\newcommand{\Xop}[1][]{{\qo[#1]{X}}}
\newcommand{\Yop}[1][]{{\qo[#1]{Y}}}

\newcommand{\ad}[1][]{{\aop[#1]\ct}}
\newcommand{\bd}[1][]{{\bop[#1]\ct}}

\newcommand{\Bd}[1][]{{\Bop[#1]\ct}}
\newcommand{\ada}[1][]{{\ad[#1]\aop[#1]}}
\newcommand{\bdb}[1][]{{\bd[#1]\bop[#1]}}

\newcommand{\f}[2][1]{\frac{#1}{#2}}

\newcommand{\fr}[2]{\frac{#1}{#2}}

\newcommand{\ei}[1]{\ee^{\ii{#1}}}
\newcommand{\emi}[1]{\ee^{-\ii{#1}}}
\newcommand{\eit}[1]{\ee^{\ii {#1}t}}

\newcommand{\fref}[1]{fig. \ref{#1}}
\newcommand{\Fref}[1]{Fig. \ref{#1}}
\newcommand{\tref}[1]{table \ref{#1}}

\newcommand{\cref}[1]{chapter \ref{#1}}
\newcommand{\Cref}[1]{Chapter \ref{#1}}
\newcommand{\sref}[1]{section \ref{#1}}

\newcommand{\MatCont}{\textsc{MatCont }}
\begin{document}

\title{Synchronization of many nano-mechanical resonators coupled via a common cavity field}

\author{C.~A.~Holmes,  C.~P.~Meaney and G.~J.~Milburn}
\affiliation{Centre for Engineered Quantum Systems, School of Mathematical and Physical Sciences, The University of Queensland, St Lucia, QLD 4072, Australia}

\begin{abstract}
Using amplitude equations, we show that groups of identical nano-mechanical resonators, interacting with a common mode of a cavity microwave field,  synchronize to form a single mechanical mode which couples to the cavity with a strength dependent on the square sum of the individual mechanical-microwave couplings. Classically this system is dominated by periodic behaviour which, when analyzed using amplitude equations, can be shown to exhibit multi-stability. In contrast groups of sufficiently dissimilar nano-mechanical oscillators may lose synchronization
and oscillate out of phase at significantly higher amplitudes. Further the method by which synchronization is lost resembles that for large amplitude forcing which is not of the Kuramoto form.   
\end{abstract}

%\ams{34C15; 37G15}

\maketitle

\section{Introduction}\label{s:MO_i}

Synchronisation of coupled oscillators arise in many different contexts in biology, chemistry and engineering \cite{guckenheimer:1983,pikovsky:2003}. Such systems show surprising emergent behaviour and can be used to encode and process information \cite{orosz:2009}.  In this paper we show how synchronisation can arise in arrays of nano-mechanical resonators interacting via a common electromagnetic field mode. Recent progress in opto-mechanical and nano-mechanical systems now enables very high frequency mechanical resonators to be coupled strongly to one or more modes of the electromagnetic field in a resonant cavity \cite{kippenberg:2008, milburn:2008}. This is largely driven by a desire to explore the deep quantum domain in which the mechanical resonator is prepared at or near its vibrational ground state \cite{rocheleau:2010,oconnell:2010}. As the coupling is essentially nonlinear, the resulting classical dynamics can be complex and must be thoroughly understood if one is to make sense of the quantum phenomenon.

The common feature in these systems is the so-called `radiation pressure coupling', whereby the displacement of each mechanical resonator independently changes the resonance frequency of a common electromagnetic cavity field by an amount proportional to the displacement of each mechanical resonator. This means that there is an effective conservative force acting on each mechanical resonator proportional to the circulating power in the electromagnetic cavity. If the cavity is externally driven, this interaction mediates an indirect all-to-all coupling between each of the mechanical resonators that is highly nonlinear.

If the oscillators are identical, a collective variable can be used to understand the dynamics. In this paper each of the oscillators is a bulk flexural vibrational mode of a mechanical resonator. The resulting set of equations is similar to that considered by Marquardt et al \cite{marquardt:2006}, who found that multi-stability was an important feature of the dynamics for small mechanical damping. Here we are able to derive amplitude equations for the collective variables and use these to map out regions of multi-stable behaviour in the system.

For nonidentical phase oscillators Kuramoto \cite{kuramoto:1975} used a collective variable (Kuramoto's order parameter) to characterise the synchronisation between the oscillators. More recently the collective dynamics of opto-mechanical arrays has been described by Heinrich et al. \cite{heinrich:2010}, who give some results on synchronisation based on a phase model related to Kuramoto's model. Like our model, this paper is based on the radiation pressure coupling between the field and mechanical elements. Unlike our model, the mechanical resonators in Heinrich et al. interact with a local electromagnetic field mode and are directly coupled by elastic forces. The more complex coupling in our model results in a different mechanism for the loss of synchronisation which typically occurs for large amplitude forcing not small amplitude forcing as occurs in the model of Heinrich et al \cite{heinrich:2010}. Nevertheless we are able to give specific results on synchronisation for two and three mechanical resonators interacting via a common cavity mode and relate these to the behaviour of a collective variable, which is related to the cavity field amplitude. 
 
%Much of the previous work on synchronised nonlinear oscillators is based on a direct, usually nearest-neighbour, interaction between the individual oscillators. Synchronisation in coupled micro-electromechanical systems (MEMS) has been described \cite{hoppensteadt:2002} and observed \cite{zalalutdinov:2003}.  In recent study on micro-mechanical optical cavities, which reduces to a  very similar set of equations to those considered here, Marquardt et al. \cite{marquardt:2006} found that multi-stability was an important feature of the dynamics for small mechanical damping. More recently, Henrich et al \cite{heinrich:2010} gave some results on synchronisation in similar system involving a large opto-mechanical array. Like our model, this paper is based on the radiation pressure coupling between the field and mechanical elements but, unlike our model, the mechanical resonators in Heinrich et al. interact with a local electromagnetic field mode and are directly coupled by elastic forces.
Much of the previous work on synchronised nonlinear oscillators is based on a direct, usually nearest-neighbour, interaction between the individual oscillators and amplitude equations have been used successfully to analyse the dynamics of such models \cite{lifshitz:2008}.  We show that amplitude equation methods can also be applied to understand the dynamics of the more complex all-to-all coupling that occurs in our model. Synchronisation in coupled micro-electromechanical systems (MEMS) has been described \cite{hoppensteadt:2002} and observed \cite{zalalutdinov:2003}. 

There are at least four kinds of physical implementations of the system discussed here. Firstly, in circuit QED, a coplanar microwave cavity contains the electric field which forms the common field mode. Nano-mechanical resonators can then be placed so as to form one plate of a capacitor with the central conductor of the microwave cavity thereby modulating the microwave cavity frequency \cite{teufel:2008b}. Secondly, at optical rather than microwave frequencies, an opto-mechanical system can be formed by placing micro-mechanical dielectric membranes inside the optical cavity \cite{thompson:2008}. Thirdly, a toroidal optical whispering gallery mode (WGM) cavity is manufactured on a tapered platform raised off a substrate \cite{anetsberger:2009}. The mechanical vibrations of the toroid modulates the frequency of the WGM. Finally, opto-mechanical phononic crystals can be fabricated which are simultaneously photonic crystal lattices to produce localised optical and mechanical modes \cite{chan:2009,lin:2010}.

In the bulk of this paper we will consider a nano-mechanical system where a single mode of a superconducting microwave resonator is coupled to the displacements of $N$ nano-mechanical resonators. However, the dynamical model we derive in this section is applicable to the other physical implementations in different experimental contexts. In general our model applies to a system of $N+1$ oscillators: $N$ single flexural modes of independent mechanical resonators whose displacements are coupled to a common single electromagnetic field mode, also modelled as a single simple harmonic oscillator.  The coupling between each mechanical resonator and the microwave field in the cavity is capacitive and results in a frequency shift of the cavity resonance frequency that, to lowest order, is proportional to the displacement of the mechanical resonator. This results in a force on each mechanical resonator that is proportional to the intensity of the microwave field in the cavity.  This is often called radiation pressure coupling \cite{walls:2008}. A schematic of this system is given in \Fref{f:MO_phys_setup}.

\fig[4]{f:MO_phys_setup}{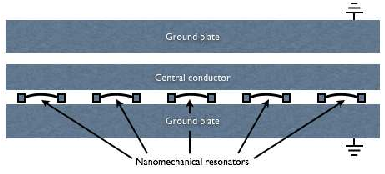}{
 A schematic of the nano-electromechanical system under consideration. A superconducting microwave cavity of frequency $\om_c$ mediates a coupling between $N$ nano-mechanical resonators capacitively coupled to it. The $i$th nano-mechanical resonator has resonant frequency $\om_i$ and microwave-mechanical coupling strength $g_i$. The microwave cavity is driven by a linear drive of amplitude $\ep$ at a detuning from the cavity of $\de$.
}{A schematic of the nano-electromechanical system under consideration}

We model the dynamics of the microwave field in the co planar transmission line by a lumped circuit $LC$ electrical resonator and the dynamics of each mechanical resonator is modelled as a single simple harmonic oscillator. The Hamiltonian for a single nanomechanical resonator interacting with the microwave field is 
\eq{
 \ham = \f[\iPh^2]{2L} + \f[Q^2]{2C(q)} + v(t)Q + \f[p^2]{2m} + \f[m\om^2]{2}q^2 \,,
}
where the first term is the inductive energy with the $\iPh$ the flux through the equivalent inductor with inductance $L$. The second term is the charging energy with $Q$ the charge on the equivalent capacitor with capacitance $C(q)$, which varies with the displacement of the mechanical element.  The third term represents the potential energy due to an external AC bias voltage of the equivalent circuit resonator. The fourth term is the kinetic energy of the mechanical resonator of effective mass $m$ and the last term is the elastic potential energy of a single flexural mode of the  mechanical resonator with $\om$.  As the displacement is small compared to the equilibrium distance between the mechanical resonator and the central conductor of the microwave cavity we can expand $C(q)$ to linear order in $q$ around the equilibrium displacement $q_0$ to get an effective Hamiltonian
\eq{
 \ham = \f[\iPh^2]{2L} + \f[Q^2]{2C_0} + \f[p^2]{2m} + m\om^2q^2 + AQ^2q + v(t)Q \,,
}
where $C_0=C(q_0)$ and $A=-\f{2}\left .\D{C(q)}{q}\right |_{q=q_0}$. The classical Hamilton's equations are
\eq{
 \Dt{\iPh} & = \f[Q]{C_0}+2AQq+v(t) \,, \\
 \Dt{Q}    & = -\f[\Ph]{L} \,, \\
 \Dt{q}    & = \f[p]{m} \,, \\
 \Dt{p}    & = -m\om^2 q - AQ^2 \,.
}
When $A=0$, the circuit equations of motion describe simple harmonic oscillation at the frequency 
\eq{
 \om_c = \f{\sqrt{LC_0}} \,.
}
It is convenient at this point to define dimensionless canonical variables. We do this by first fixing two energy scales, one for the circuit degrees of freedom $E_c$ and one for the mechanical degrees of freedom, $E_m$. The dimensionless canonical variables, $(x_c,y_c)$ for the circuit and $(x,y)$ for the mechanics,  are then defined by 
\eq{
 x_c & = \f[\Ph]{\sqrt{2E_cL}} \,, \\
 y_c & = \f[Q]{\sqrt{2E_cC_0}} \,, \\
 x   & = \f[q]{\sqrt{\f[2E_m]{m\om^2}}} \,, \\
 y   & = \f[p]{\sqrt{2mE_m}} \,.
}
We now anticipate an eventual quantum mechanical treatment and set $E_c=\hbar\om_c$, $E_m=\hbar\om$. The appearance of $\hbar$ at this stage does not signify anything more than a convenient conversion factor between energy and frequency. We also define a complex amplitude for the circuit degrees of freedom as 
\eq{
 \al = x_c + \ii y_c \,,
}
in terms of which we can write the Hamilton's equations of motion as 
\eq{
 \Dt{\al} & = -\ii\om_c\al - \ii g\<{\al-\al^*}x + {\cal E}(t) \,, \\
 \Dt{x}   & = \om y \,, \\
 \Dt{y}   & = -\om x - gy_c^2 \,,
}
where 
\eq{
 g           & = \f[\sqrt{2}\,AC_0E_c]{\sqrt{mE_m}} \,, \\
 {\cal E}(t) & = \f[v(t)]{\sqrt{2E_cL}} \,.
}
We now assume that the circuit is harmonically driven and set
\eq{
 {\cal E}(t) = {\cal E}_0\sin\om_D t
}
and define the rotating variable $\bar{\al}=\al\eit{\om_D}$ (equivalent to going to the interaction picture in the quantum description). If we then drop rapidly rotating terms (compared to the time scale of observations), the equations of motion may be approximated by
\eq{
 \Dt{\bar{\al}} & = -\ii\de\bar{\al} - \ii g\bar{\al}x - \ii\ep \,, \\
 \Dt{x}         & = \om y \,, \\
 \Dt{y}         & = -\om x - \f[g]{2}\abs{\bar{\al}}^2 \,,
}
where $\ep=-\f[{\cal E}_0]{2}$, and the detuning $\de=\om_c-\om_D$. Noting the the time averaged energy of the energy in an LC circuit is proportional to $\abs{\al}^2$ we see that the effective coupling between  the microwave field and the mechanical resonators is described by the effective Hamiltonian $g\abs{\al}^2 x$. This form of coupling is often termed radiation-pressure coupling \cite{walls:2008} being proportional to the circulating power in the cavity field. 

We include dissipation of both the microwave field mode and the nano-mechanical resonators using the quantum mechanical master equation to incorporate fluctuations correctly. We derive this in section \ref{s:quantum}. There we show that in the classical description, the systematic effect of damping (i.e. ignoring fluctuations) change the Hamilton equations to
\eq{
 \Dt{\al} & = -\ii\de\al - \ii g\al x - \ii\ep - \ka\al \,, \\
 \Dt{x}   & = \om y - \ga x \,, \\
 \Dt{y}   & = -\om x - \f[g]{2}\abs{\al}^2 - \ga y \,,
}
where $\ka,\ga$ are the energy decay rates for the electrical and mechanical energy respectively and we have dropped the bar as from this point on we simply take it for granted that we are working with the rotating variables for the cavity field.

In this paper we are interested in the dynamics of $N$ mechanical resonators interacting with a single mode of the microwave field in the circuit. Assuming a coupling of the form $\sum_i g_ix_i \abs{\al}^2$, we see that the equations of motion may be expressed in terms of collective variables. 
\eq[e:origdes]{
 \Dt{\al} & = -\ii\de\al - \ii\ep - \ii\al\sum_{i=1}^MN_iX_i - \ka\al \,, \\
 \Dt{X_i} & = \om_iY_i - \ga_iX_i \,, \\
 \Dt{Y_i} & = -\om_iX_i - \f[G_i]{2}\abs{\al}^2 - \ga_iY_i \,,
}
where $X_i$, $Y_i$ and $G_i$ are the $M$ collective combinations
\eq{
 X_i & = \f{N_i}\sum_{j\in S_i}g_jx_j \,, \\
 Y_i & = \f{N_i}\sum_{j\in S_i}g_jy_j \,, \\
 G_i & = \f{N_i}\sum_{j\in S_i}g_j^2 \,,
}
and $S_i$ are collections of $N_i$ identical nano-mechanical oscillators with individual classical positions and momenta $x_j$ and $y_j$ respectively. We note that the other experimental contexts mentioned in this introduction can also be described by the same differential equations \eqref{e:origdes}. For example, multiple opto-mechanical membranes in an optical cavity with are described by these equations with different resonant frequencies and coupling strengths \cite{thompson:2008}. We give a list of the achievable experimental values for various experiments in \tref{t:MO_exp1} in \sref{s:MO_aexp}.

%{\bf Delete this paragraph because it is obvious? else change it to be classical} At this point we note that both $\<{\ep,\aop}\mapsto\<{\ep\ei{\ps},\aop\ei{\ps}}$ and $\<{g_i,\bop[i]}\mapsto\<{-g_i,-\bop[i]}$ are symmetries of the system, where $\ps\in\as{R}$. Thus without loss of generality, we can assume that $\ep\in\as{R}$, $\ep\ge0$, and $g_i\ge0$. Equivalently, we can make the redefinition mappings $\ep\mapsto\abs{\ep}$, $g_i\mapsto\abs{g_i}$, $\aop\mapsto\sgn\<{\ep}\aop$, and $\bop[i]\mapsto\sgn\<{g_i}\bop[i]$, where we mean $\sgn\<{z}=1$ if $z=0$ and $\sgn\<{z}=\f[z]{\abs{z}}=\ei{\angle z}$ otherwise (the condition being to avoid the potential singularity). The commutators are preserved, $\C{\aop}{\ad}\mapsto\C{\aop}{\ad}=1$, $\C{\bop[i]}{\bd[j]}\mapsto\C{\bop[i]}{\bd[j]}=\de_{i,j}$. For the remainder of the document we will use these simplifications. We can invert the mapping to write expressions in terms of the original parameters if desired.

In the following section, we present a detailed analysis of the steady state structure of the nonlinear semi-classical system, including local and global bifurcations. Since the behaviour is dominated by oscillatory motion, amplitude equations are derived from which we can obtain specific results about the existence and stability of periodic orbits. It is then a simple step to derive coupled amplitude equations for the case where the mechanical oscillators are not identical and we analyse two and three coupled oscillator systems. In \sref{s:quantum} we give a quantum description of the many-body system, and calculate the steady state quantum noise spectra as the first stable limit cycle is approached. Finally in section \sref{s:MO_c} we summarise our results and suggest new directions for further work.

\section{Dynamics of the Classical model}\label{s:MO_sc}

Although there are regions of the parameter space where stable critical points exist, periodic motion plays a major role in the dynamics for the cases of both the identical and the nonidentical resonators. If the mechanical resonators are identical, even if their couplings are nonidentical, they will synchronize, in phase, to form a single mechanical mode. However the synchronized motion exhibits multi stable behaviour. The first two sections, below, discuss the synchronized motion of  identical mechanical resonators \eqref{e:eomr}, largely via amplitude equations. If, on the other hand, the mechanical resonators naturally oscillate at different frequencies, desynchronization can occur. To analyze this we consider the synchronization between different frequency groups. The resonators can then be attracted to  out-of-phase solutions that oscillate at much greater amplitudes. In the final section we obtain specific results, via coupled amplitude equations, for synchronization between two and three frequency groups. 

% The actual bifurcations involved in the loss of synchronization turn out to be Hopf In the second section we investigate the loss of synchronization that can occur if the resonators are nonidentical.

For all of the bifurcations that occur a scaled version of the cavity forcing $\ep$, which is tunable in an experiment, can be thought of as the bifurcation parameter. There are two time scales in the system; the amplitude decay rate $\ka$ of the common cavity mode and the decay rate of the resonators, which is an order of magnitude smaller and will be important for the derivation of the amplitude equations. The amplitude decay rate $\ka$ of the common cavity mode provides a natural time-scale and we introduce: a new time parameter $t\pr=\ka t$; re-scaled nano-mechanical variables $X_i\pr=\f[X_i]{\ka}$ and $Y_i\pr=\f[Y_i]{\ka}$; and dimensionless coupling constants $\de\pr=\f[\de]{\ka}$, $\ep\pr=\f[\ep]{\ka}$, $\om_i\pr=\f[\om_i]{\ka}$, $\ga_i\pr=\f[\ga_i]{\ka}$, $G_i\pr=\f[G_i]{\ka^2}$, and $\bar{\om_i}\pr=\sqrt{{\om_i\pr}^2+{\ga_i\pr}^2}$. This gives
\eq[e:eomr]{
 \D{\al}{t\pr}       & = -\<{1+\ii\de\pr}\al - \ii\al\sum_{i=1}^MN_iX_i\pr - \ii\ep\pr \,, \\
 \D[2]{X_i\pr}{t\pr} & = -\nm{\bar{\om_i}\pr}^2X_i\pr - \f[G_i\pr\om_i\pr]{2}\abs{\al}^2 - 2\ga_i\pr\D{X_i\pr}{t\pr} \,.
}
%In addition, $N_i$ could be removed by scaling
%\eq{
% \bar{X_i}\pr & = N_iX_i\pr \,, \\
% \bar{Y_i}\pr & = N_iY_i\pr \,, \\
% \bar{G_i}\pr & = N_iG_i\pr \,.
%}
If the uncoupled mechanical resonators are identical ($\om_1\pr=\om\pr$, $\ga_1\pr=\ga\pr$  $\imp\bar{\om_1}\pr=\bar{\om}\pr$), then  the oscillators synchronize.  This is a natural consequence of linear damping and the fact that each oscillator experiences the same forcing. Consider $u=X_i\pr-X_j\pr$ then $u =0$ is a stable solution of its equation of motion: 
\eq{
 \D[2]{u}{t\pr} & = -\nm{\bar{\om}\pr}^2u - 2\ga\pr\D{u}{t\pr} \,,
}
provided $\ga\pr>0$.

The synchronized motion can then be represented in  collective variables \eqref{e:eomr} which, suppressing the use of primes, gives the following
\eq{
 \D{\al}{t}     & = -\<{1+\ii\de}\al - \ii\al NX - \ii\ep \,, \\
 \D[2]{X}{t} & = -\nm{\bar{\om}}^2X - \f[G\om]{2}\abs{\al}^2 - 2\ga\D{X}{t} \,.
}
For the remainder of this paper we suppress the uses of primes in the notation, and remind the reader that all couplings are now dimensionless with the cavity decay rate determining the natural time-scale of the system.

From a dynamical point of view $\ep\sqrt{NG}$ acts as one parameter and in fact both $N$ and $G$ could be removed by scaling
\eq{
 \bar{X}   & = NX \,, \\
 \bar{\al} & = \al\sqrt{NG} \,, \\
 \bar{\ep} & = \ep\sqrt{NG} \,.
}
So if the number of resonators is increased, smaller values of the driving are necessary to achieve the same effect.

\subsection{Critical points, bifurcations and stability}

\label{ss:bifs}

Without forcing, $\ep=0$, the origin is a stable critical point. As the $\ep$ is increased from zero the critical point moves away from the origin, its position given by the single real root of the cubic 
\eq{
 2\bar{\om}^2X_0\<{1+\<{\de+NX_0}^2} + G\om\ep^2 = 0 \,,
} 
where $\al_0 = -\f[\ii\ep]{1 +\ii\<{\de+NX_0}}$. However it loses stability on a 
 Hopf bifurcation, creating a periodic orbit, for both $\de>0$ (red detuning) and $\de<0$ (blue detuning) provided $\ga>0$ and small. The dynamics of this periodic motion is the subject of the next section. 

\begin{figure}[!htbp]\begin{center}
\includegraphics[scale=0.7]{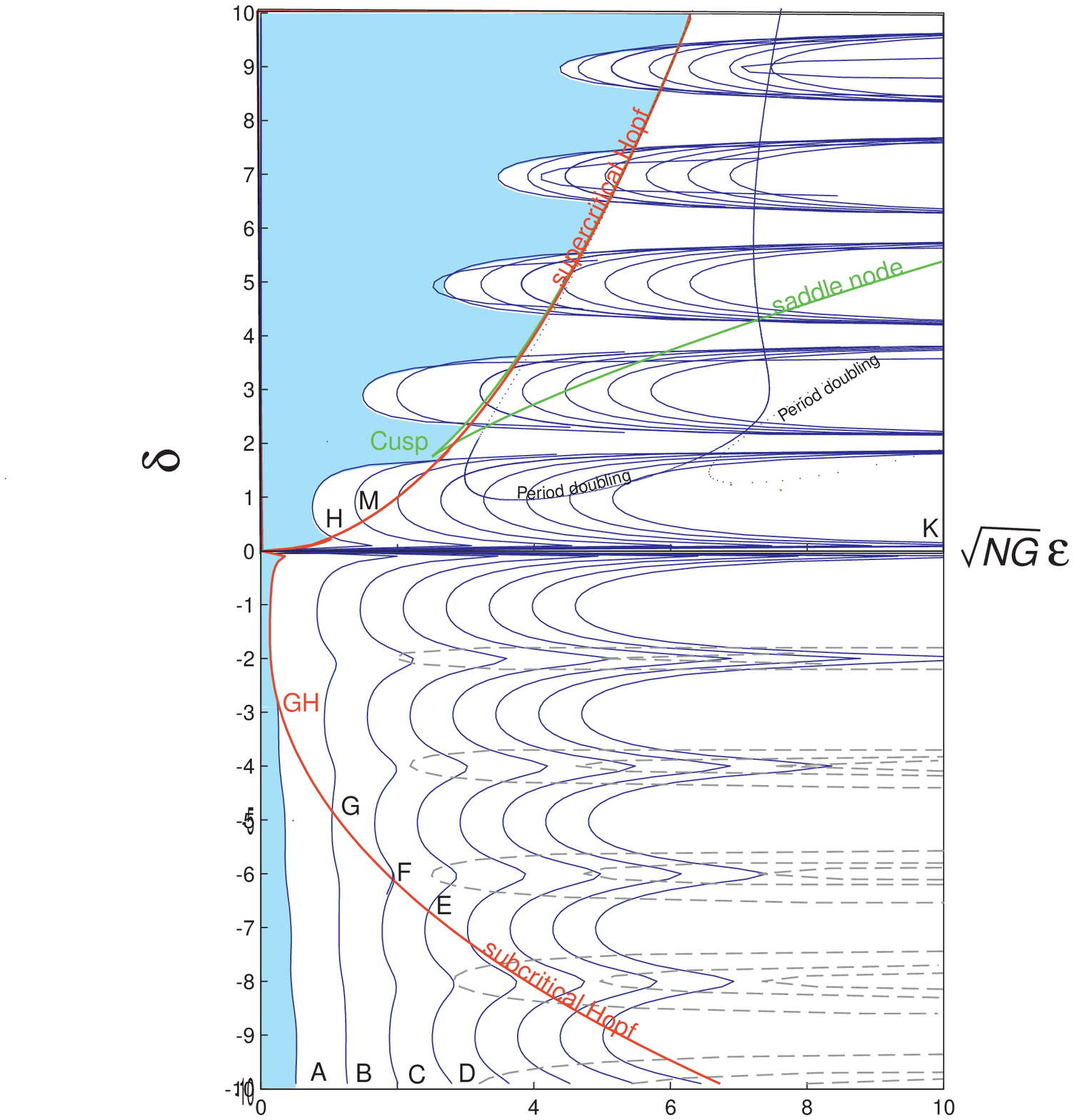}
\caption{The bifurcation diagram  for $\om=2$ and $\ga=0.001$.
In the shaded region there are no periodic orbits and there is one stable critical point. The Hopf bifurcation curve, which is in red, provides a partial boundary of this region. At the generalized Hopf GH (which is at  $\de=\sqrt{7}$ for $\om=2$) the Hopf bifurcation changes from super to sub critical. For $\de<\sqrt{7}$ the Hopf bifurcation is sub critical  and periodic orbits exist to the left of the Hopf curve. Also for $0<\de<\sqrt{3}$ there are regions where periodic orbits exist to the left of the Hopf curve. The blue curves A-GH, BGK, CFK , DEK, KHCusp, KMcusp etc are saddle node bifurcations of periodic orbits creating a stable and an unstable periodic orbit existing to their right. (Only the first 8 are shown.) The lozenge like  dashed curves  are also saddle node bifurcations of periodic orbits, this time destroying a stable and an unstable periodic orbit. (Once again only a sample are shown.)  In region ABG(GH) and HM Cusp there is one stable critical point and a pair of periodic orbits with opposite stability. In region G(GH)HK there is one unstable critical point and one stable periodic orbit. In region BCFG and the region to the left of M Cusp there is one stable critical point and two pairs of periodic orbits with opposite stability. In region FGK there is one unstable critical point and two stable periodic orbits and one unstable periodic orbit. In region CDEF there is one stable critical point and three stable and three pairs of periodic orbits with opposite stability  etc.}
\label{f:figure2}
\end{center}\end{figure}

For $\de>0$ (red detuning) the Hopf curve  is a perturbation of that from  the $\ga=0$ case where $\sqrt{NG}\;\ep = \sqrt{2\de\bar{\om}}$. To first order in $\ga$ it is given by
\eq{
 \ep = \ep_H\<{\om,\de,\ga,NG} = \sqrt{\f[2\om]{NG}\<{\de + \ga\f[\<{1+\om^2}^2]{2\de\om^2}}} \,.
}
For $\de<0$ (blue detuning) $\ep$ is order $\sqrt{\ga}$:
\eq{
 \ep = \ep_H\<{\om,\de,\ga,NG} = \sqrt{\f[\ga\<{1+\de^2}\<{\<{\de^2-\om^2+1}^2+4\om^2}]{-\de NG\om}} \,.
}
The Hopf bifurcation is subcritical   for $\de < -\sqrt{\f[8\om^2+3]{5}}$ (blue detuning), where  periodic orbits can exist for $\ep < \ep_H\<{\om,\de,\ga,NG}$.
In fact many stable limit cycles can exist for some parameter values because  of the presence of saddle node bifurcations of limit cycles each creating a stable and unstable pair of limit cycles. This leads to multi stable behaviour that has been noticed elsewhere \cite{marquardt:2006,dorsel:1983} for similar systems. These bifurcations are shown in \fref{f:figure2} for $\om=2$ and $\ga=0.001$. The limit cycle bifurcations were produced using using the amplitude equations described in the next section, however similar results can be produced by following the limit cycles numerically using the package \MatCont \cite{matcont}. For $\de<0$ (blue detuning) eight of the saddle node bifurcations of limit cycles are shown indicating regions where there are 1-8 pairs of stable and unstable periodic orbits. See the caption for specific details. \MatCont indicates the situation is dynamically more complicated for $\de>0$ (red detuning) involving period doubling and regions of chaos.

Although most of this paper is devoted to the case of blue detuning, where $\de<0$,  it is worth mentioning that for $\de>\sqrt{3}$ there is a region where three critical points exist given by the roots of the cubic given above. This triangular shaped region
\eq{
 \f[2\bar{\om}\<{2\de+\sqrt{\de^2-3}}]{3\sqrt{NG\om}\sqrt{\de+\sqrt{\de^2-3}}} = \ep_{sn+} < \ep < \ep_{sn-} = \f[2\bar{\om}\<{2\de-\sqrt{\de^2-3}}]{3\sqrt{NG\om}\sqrt{\de- \sqrt{\de^2-3}}} \,,
}
is bounded by ($\ep=\ep_{sn\pm}$) saddle node bifurcations, shown as green lines in \fref{f:figure2}. These intersect in a cusp bifurcation at $\de = \sqrt{3}$ and $\ep \sqrt{NG\om} = \f[4\bar{\om}]{\sqrt{3}}$. 

\begin{figure}[!htbp]\begin{center}
\includegraphics[scale=0.5]{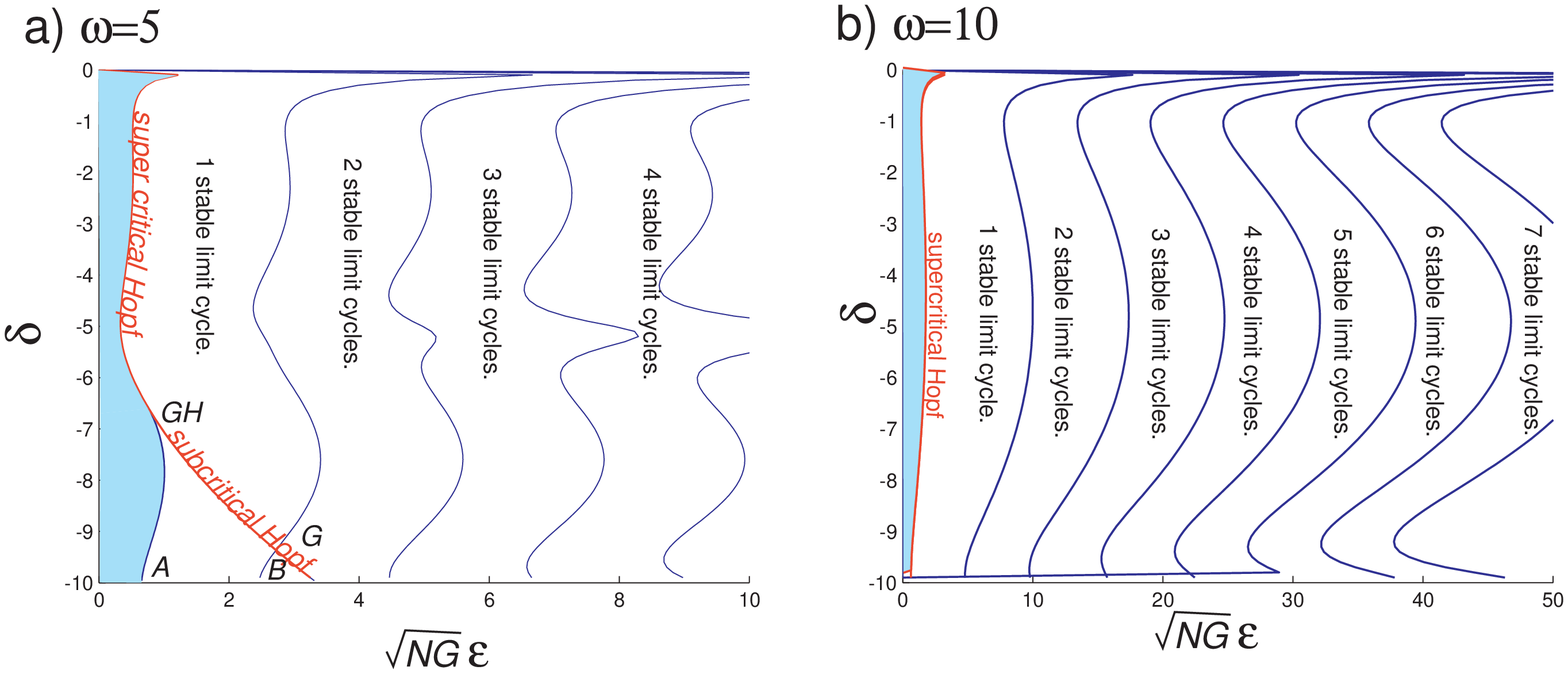}
\caption{Bifurcation diagrams a) $\om=5$ and b) $\om=10$ and $\ga=0.001$ with $\de<0$ (blue detuning) showing the Hopf bifurcation (red) and saddle node bifurcations of periodic orbits.   The labeling in 3a) is similar to \fref{f:figure2}. For instance in the region ABG(GH) there is one stable critical point and a pair of periodic orbits with opposite stability.}
\label{f:figure3}
\end{center}\end{figure}

The case $\om=2$ is relevant for the experiments described in \cite{marquardt:2006,heinrich:2010}. However for $\om>2$, as in \cite{teufel:2008,sulkko:2010}, there is no qualitative change in the bifurcation diagram, although the generalized Hopf bifurcation  ($\de = -\sqrt{\f[8\om^2+3]{5}}$) occurs for larger values of $\abs{\de}$. \Fref{f:figure3} shows the corresponding situation for a) $\om=5$ and b) $\om=10$ and $\ga=0.001$ with $\de<0$ (blue detuning). Multistable behaviour due to the presence of limit cycles stacked above each other remains an important feature (see also \fref{f:figure4}).  

% For $\om=10$ the Hopf bifurcation remains supercritical for $-10<\de<0$. Note that the scale is different in 3a) and 3b).  For larger values of $\om$ the periodic orbits have significantly larger amplitude and exist for larger values of $\sqrt{NG}\ep$.

\subsection{Amplitude Equations and multi-stability for blue detuning ($\de<0$)}
 
Periodic orbits and  multiple periodic orbits can exist, if the weakly forced oscillators are sufficiently weakly damped. This multi stable behaviour, resulting from the play off between weak damping and cavity forcing,  has been noted elsewhere \cite{marquardt:2006,heinrich:2010,mueller:2008}. Here we explore it in more depth  using amplitude equations.  

 The method relies on defining a slow time which is proportional to the damping rate of the resonators: $(\ta = \ga t)$ and on assuming that the forcing is  on the order of the square root of the damping: $\ep = \sqrt{\ga}\,\bar{\ep}$. Then the cavity amplitude is naturally of the same order  as the forcing and we can obtain equations for the slowly varying amplitude $A(\ta)$. Let
\eq{
 X = X_0+\<{A(\ta)\eit{\bar{\om}} + \txt{c.c.}} = X_0 + 2\abs{A(\ta)}\cos\<{\bar{\om}t + \thet} \,,
}
where $X_0$ is the critical point of the system given in the previous section, 
which is $O(\ep^2)$. Given that $\ga$ is small and both $\ep$ and $\abs{\al}$ are $ O(\sqrt{\ga})$ then $\ddot{X} + \bar{\om}^2 (X-X_0) \approx 2\ga\ii\om\D{A}{\ta}\ee^{\ii\bar{\om}t} + \mathrm{c.c.}$ \cite{glendinning:1994}. The cavity forcing $(\f[G\om]{2}\abs{\al}^2)$ can then be written as a sum of products of Bessel functions. To see this substitute $X = X_0 + 2\abs{A}\cos\<{\bar{\om}t + \thet}$ into the cavity equation;
\eq{
 \Dt{\al} = -\<{1 + \ii\<{\de+N X_{0} + 2N\abs{A}\cos\<{\bar{\om} t +\thet}}}\al - \ii\ep \,.
}
Then if
\eq{
 \al = \ei{\ps(t)} \sum_m B_m \ei{m\om t}
}
it follows that
\eq{
 \dot{\al} = \ii\dot{\ps}(t) \al + \ei{\ps(t)}\sum_m\ii m\om B_m\ei{m\om t} \,,
}
and using the Jacobi Anger expansion \cite{marquardt:2006} $\sum_{n=-\infty}^{\infty}\ii^nJ_n(z)\ei{n\thet}=\ei{z\cos\thet}$, this can be matched to the right hand side  of the cavity equation  if
\eq{
 \dot{\ps(t)} = -\f[2N\abs{A}]{\om}\cos\<{\om t+\thet} \hspace{5mm}\txt{and}\hspace{5mm} B_m = -\fr{\ii^{m+1}\ep J_m\<{\f[2N\abs{A}]{\om}}}{\bar{\ka}+\ii m\om} \,,
}
where $\bar{\ka} = 1 + \ii\<{\de+ NX_0}$ and $J_m(x)$ are Bessel functions of the first kind. Substituting this back into
\eq{
 \ddot{X} = -\bar{\om}^2X - \f[G\om]{2}\abs{\al}^2 - 2\dot{X}
}
gives an amplitude equation for the oscillation in terms of sums of pairs of Bessel functions,
\eq[e:ampeqs]{
 \D{A}{\ta} = -A -\f[\ii G\ep^2\ei{\thet}]{4}\sum_{m=-\infty}^{\infty}\fr{J_m\<{\f[2N\abs{A}]{\om}}J_{m+1}\<{\f[2N\abs{A}]{\om}}}{\<{\bar{\ka}+\ii\<{m+1}\om}\<{\bar{\ka}^*-\ii m\om}} \,.
}
Identical mechanical resonators synchronize to oscillate with amplitude $A(\ta)$ given by this equation. 

\begin{figure}[!htbp]\begin{center}
\includegraphics[scale=1]{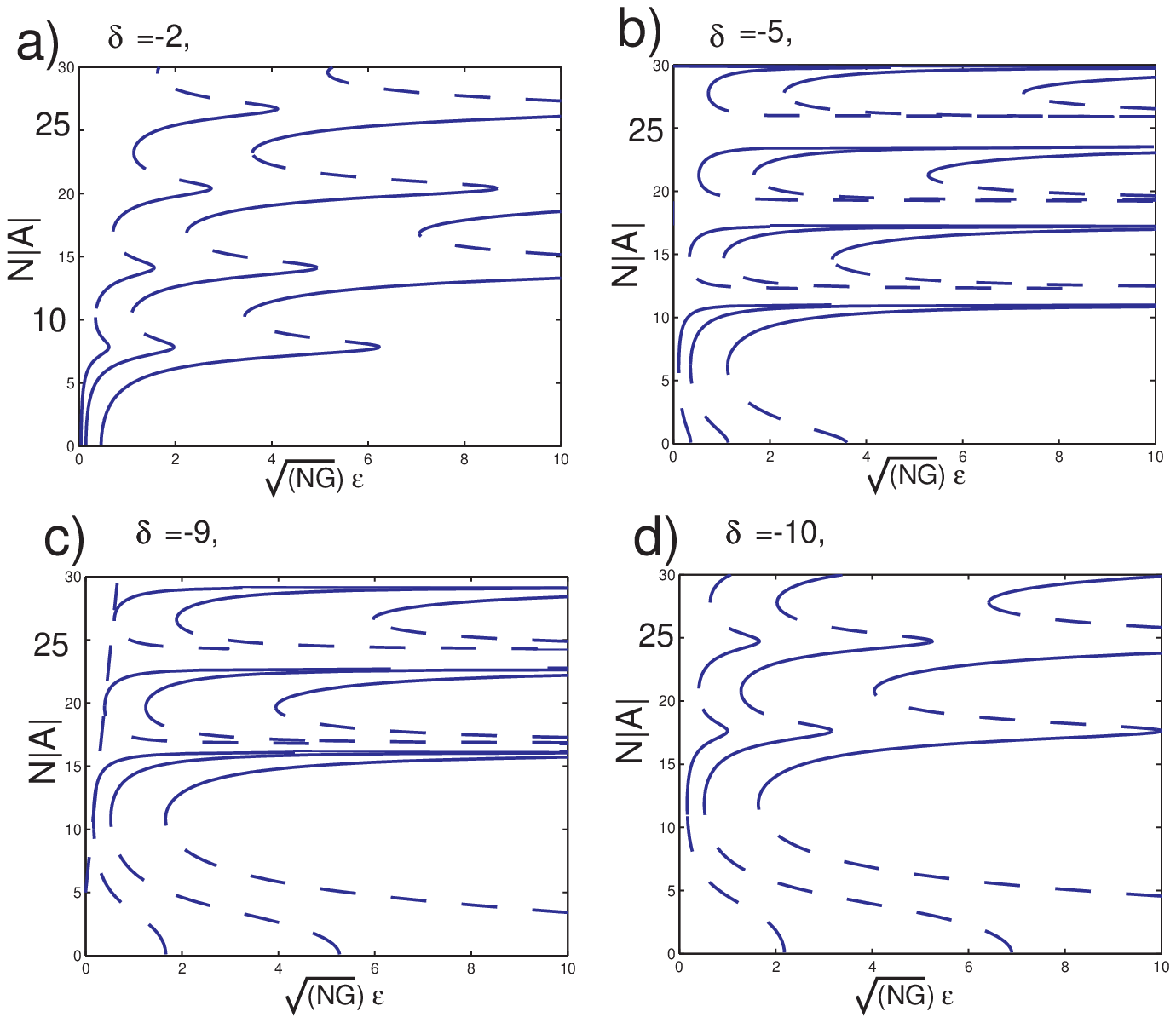}
\caption{The amplitudes ($N\abs{A}=Nr$) of the periodic orbits of the system calculated from the amplitude equations as a function of $\sqrt{NG}\;\ep$  for $\om=2$ and $\ga=0.01,\,0.001,\,0.0001$ and various values of $\de$. In a) $\de=-2$, b) $\de=-5.$, c) $\de=-9$, d) $\de=-10$. The unstable periodic orbits are given by dashed lines and the stable one are given by solid lines.}
\label{f:figure4}
\end{center}\end{figure}

%Since $J_{-m}(x) = (-1)^{m}J_m(x)$ the doubly infinite sum can be written as a sum over the positive integers;
%\eq{
%  \sum_{m=0}^{\infty}   \fr{J_m\<{\f[2N\abs{A}]{\om}}J_{m+1}\<{\f[2N\abs{A}]{\om}}(2m+1)\ii\om(\bar{\ka}-\bar{\ka}^*)}{(\bar{\ka}+\ii(m+1)\om)(\bar{\ka}^*-\ii m\om) (\bar{\ka}^*+\ii(m+1)\om)(\bar{\ka}-\ii m\om)} \,.
%}
In polar form ($A=r\ei{\thet}$) the equations become
\eq{
 \D{r}{\ta}     & = - r + G\bar{\ep}^2\sum_{m=0}^{\infty} a_{mr}\<{\bar{\de},\om} J_m\<{\f[2Nr]{\om}}J_{m+1}\<{\f[2Nr]{\om}} \,, \\
 \D{\thet}{\ta} & = +\f[G\bar{\ep}^2]{r}\sum_{m=0}^{\infty} a_{mi}\<{\bar{\de},\om} J_m\<{\f[2Nr]{\om}}J_{m+1}\<{\f[2Nr]{\om}} \,,
}
where
\eq{
 a_{mr}\<{\bar{\de},\om} & = \f[\bar{\de}\om^2\<{2m+1}\<{1+\bar{\de}^2 +\om^2 m\<{m+1}}]{\<{\<{1+\bar{\de}^2-m^2\om^2}^2+4m^2\om^2}\<{\<{1+\bar{\de}^2-\<{m+1}^2\om^2}^2+4\<{m+1}^2\om^2}} \,, \\
 a_{mi}\<{\bar{\de},\om} & = \f[\bar{\de}\om\<{2m+1}\<{\<{1+\bar{\de}^2 -m^2\om^2}\<{1+\bar{\de}^2 -\<{m+1}^2\om^2}+4 m\<{m+1}\om^2}]{2\<{\<{1+\bar{\de}^2-m^2\om^2}^2+4m^2\om^2}\<{\<{1+\bar{\de}^2-\<{m+1}^2\om^2}^2+4\<{m+1}^2\om^2}} \,,
}
and $\bar{\de}=\de + X_0$. For $\de<0$ (blue detuning) then $X_0\approx\f[\ga\<{\<{\de^2-\om^2+\ka^2}^2+4\ka^2\om^2}]{2\ka\om^2\de}$. Since each term in the sum has $\abs{A}$ as a factor, the amplitude equation may be rewritten as
\eq{
 \D{A}{\ta} = -  A +G\bar{\ep}^2 N A F(N \abs{A},\,\om,\,\de) \,,
}
%and in polar form as
%\eq{
% \D{r}{\ta}     & = - r +G\bar{\ep}^2 Nr F_r\<{Nr,\om,\de} \,, \\
 %\D{\thet}{\ta} & = NG\bar{\ep}^2 F_i\<{Nr,\om,\de} \,,
%}
where  $F\<{Nr,\om,\de}$ is a complex function.
% and the subscripts denote its real and imaginary parts. At $r=0$,
%\eq{
% F_r(0,\,\om,\,\de) = \f[a_{0r}]{\om} =\f[\bar{\de}\om]{(1+\bar{\de}^2)(((1+\bar{\de}^2-\om^2)^2+ 4\om^2)} \,.
%}
The conditions for the Hopf bifurcation, given in section \ref{ss:bifs}, can be obtained by setting $\D{r}{\ta}=0$ in the linearized radial equation,
%\eq{
% NG\bar{\ep}^2 = \f[1]{F_r(0,\,\om,\,\de)}=\f[\om]{a_{0r}} \,.
%}

Since $\thet$ does not appear in the equation for $r$, the periodic orbits of the system are given by
\eq{
 F_r\<{Nr,\om,\de} = \f{r}\sum_{m=0,\infty} a_{mr}\<{\bar{\de},\om} J_m\<{\f[2Nr]{\om}}J_{m+1}\<{\f[2Nr]{\om}} =  \f{NG\bar{\ep}^2 }\,.
}
These curves are plotted in \fref{f:figure4} for $\om=2$ and $\ga=0.01,\,0.001,\,0.0001$ and various values of $\de$. Corresponding to these oscillations, the cavity field amplitude oscillates with frequency $\bar{\om} + F_i\<{Nr,\om,\de}$ and amplitude $\ep\sqrt{2Nr\abs{F}}$:
\eq{
 \<{\mbox{Leading oscillatory term in }\abs{\al}^2} = 2Nr\ep^2\abs{F\<{Nr,\om,\de}}\cos\<{\<{\bar{\om} + F_i\<{Nr,\om,\de}}t + \ze} \,,
}
where $\ze$ is a constant.

%It is worth pointing out the role played by $N$ at this point, because both axes in these plots are functions of $N$ in different ways. For instance the oscillations with smallest amplitude for $\de=-2$ oscillate at a radius of about 6, but for two mechanical resonators this will mean that the each resonator oscillates at a radius of about 3 and for three mechanical resonators at a radius of about 2.

\begin{figure}[!htbp]\begin{center}
\includegraphics[scale=0.38]{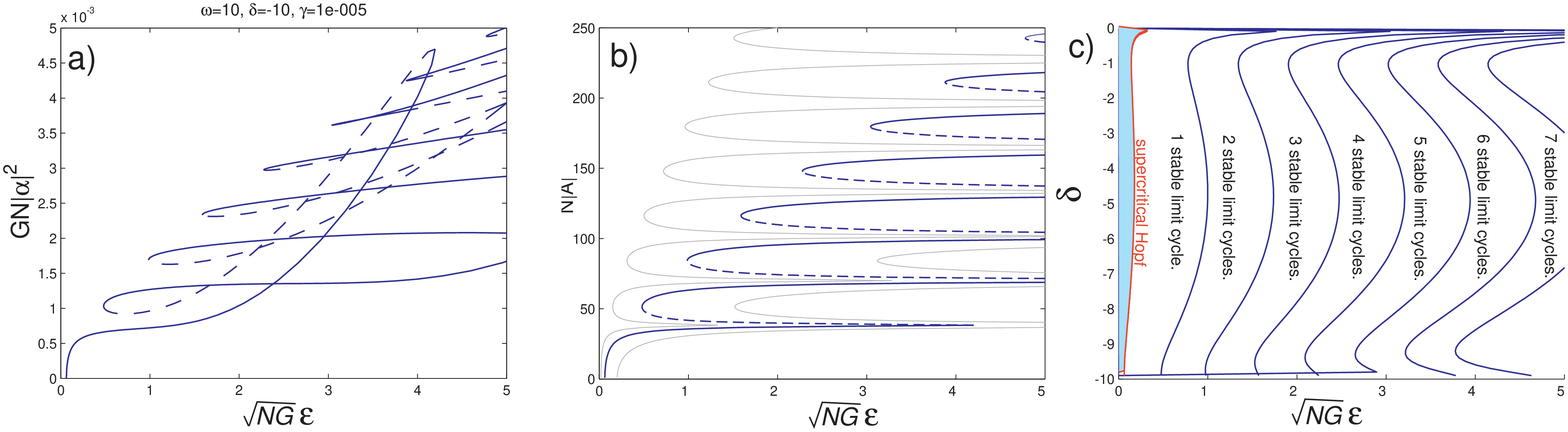}
\caption{The amplitudes  $\abs{\al}$ and  $N\abs{A}$ as a $\ep$ is increased  for $\om=10$ and $\ga=0.00001$ compared with the saddle node bifurcations that create them. Figure part a) is the cavity amplitude $\abs{\al}$, b) $N\abs{A}$ and c) the bifurcation diagram.}
\label{f:figure5}
\end{center}\end{figure}

%For some values of $N\abs{A}$ $(=Nr)$ there are neither stable nor unstable periodic orbits. To see the values of $\de$  for which there are periodic orbits for specific ranges of amplitude we have plotted the contours of the function $F_r\<{Nr,\om,\de}$ in \fref{f:figure6}.  The function surface  has mountains ranges running almost diagonally with stable periodic orbits on the `western' slopes and unstable periodic on the eastern slopes. There are multiple periodic orbits for a given value of $\de$ because the surface has multiple ranges. However not all these orbits exist for a given $\ep$ and $\de$. For instance \fref{f:figure4} a) shows that for $\de=-2$ and $\ep=1$ there are three stable periodic orbits for $\ga=0.0001$, two for $\ga=0.001$ and one for $\ga=0.01$. The contour plot also shows that there are distinct regions in $\<{Nr,\de}$ space where there are no periodic orbits  where $F_r (Nr,\,2,\de)<0$ (enclosed by red curves). 
\begin{figure}[!htbp]\begin{center}
\includegraphics[scale=1]{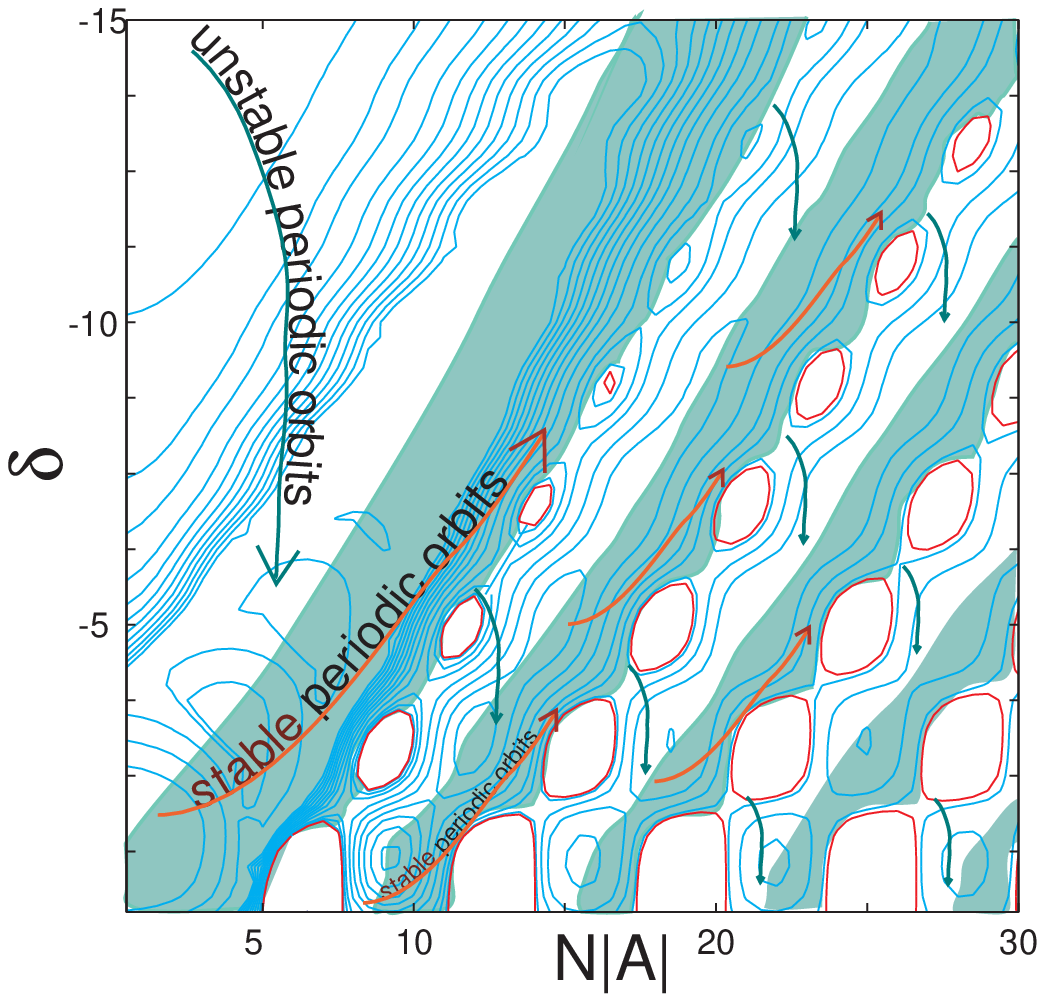}
\caption{The contours of the function $F_r\<{Nr,\om,\de}$ for $\om=2$ plotted as a function of $\<{Nr,\de}$, calculated using 10 terms in the sum of products of Bessel functions. Stable periodic orbits exist in green shaded regions. There are no periodic orbits in the regions enclosed by red lines where $F_r (N\abs{A},\,2,\de)<0$.}
\label{f:figure6}
\end{center}\end{figure}

Although the equation governing the periodic orbits is simple, the multiple ranges of the function $F_r\<{Nr,\om,\de}$, whose contours are plotted in \fref{f:figure6} for $\om=2$, result in  multi stability. Its    turning points  define  the positions of the saddle node bifurcations, which   map out the number of periodic orbits existing in parameter space, as shown in \fref{f:figure2}. (The curves shown in \fref{f:figure2} were calculated using 10 terms in the sum.) Expanded in a Taylor series as a function of $r^2$ about zero; 
\eq{
 F_r\<{Nr,\bar{\ka},\om} = F_{r0}\<{\bar{\ka},\om} + r^2F_{r1}\<{\bar{\ka},\om} + r^4F_{r2}\<{\bar{\ka},\om} + \cdots \,.
}
the linear term; $F_{r1}\<{\bar{\ka},\om}$, defines the criticality of the Hopf bifurcation. The Hopf bifurcation is super-critical, creating a stable periodic orbit, if $F_{r1}\<{\bar{\ka},\om}<0,$ which is the case here for $\ga$ small if  $\de > -\sqrt{\f[8\om^2+3]{5}}$. 

For larger values of $\om$ the oscillations occur at radii with greater values of $N\abs{A}$ $(=Nr)$. \Fref{f:figure5} compares the amplitudes $\abs{\al}$ and $N\abs{A}$ with the bifurcation diagram for $\om=10$ and $\ga=0.00001$. For instance in the experiment described in \cite{teufel:2008} the upper bound for the magnitude of epsilon implies that oscillatory behaviour occurs for $N$ on the order of 10 and multi-stable behaviour occurs for $N$ on the order of 500.

Here we will not consider the case with $\de>0$, which corresponds to red detuning, except to note that the dynamics is more complicated and deserves a separate study. While  periodic orbits, similar to those discussed here, exist there are other orbits as well, associated with the Hopf bifurcation, and many of these undergo period doubling  (see \fref{f:figure2}) to chaos. 

\subsection{$N$ nonidentical mechanical resonators and synchronization}

%To make the algebra easy scale the $\bar{x}_i=g_i\bar{x}_i$ and time with respect to $\ka$ as before.

If the frequencies and or damping of each individual mechanical resonator differs,  reduction to a single collective variables is no longer possible. However the results of the previous section can be generalized to give a set of N coupled amplitude equations. Here we consider the case where the linear frequency of the mechanical resonators are approximately the same;  $\bar{\om}_i = \om  + \ga\De\om_i$. The equations of motion, \eqref{e:eomr}, then become 
\eq{
 \Dt{\al}   & = -\<{1+\ii\de} \al- \ii\al\  \sum_iN_iX_i-\ii \ep \,, \\
 \ddot{X_i} & = -\<{\om^2+2\om\De\om_i} X_i-\f[G_i\om_i]{2}\abs{\al}^2-2\ga\dot{X_i} \,.
}
As in the previous section amplitude equations, as function of a slow time $(\ta)$, can be derived for the dominant oscillatory term \cite{lifshitz:2008}
\eq{
 X_i=X_0+\<{A_i(\ta)\ei{\om t} + \txt{c.c.}} = X_0+2\abs{A_i\<{\ta}}\cos(\om t + \thet_i) \,.
} 
Taking a sum as before and rewriting this as one oscillatory term,
\eq{
 X = \f{N}\sum_iN_iX_i=X_0+\f{N}\<{\sum_iA_i(\ta)\ei{\om t} + \txt{c.c.}} = X_0 + 2\abs{A}\cos\<{\om t+\thet} \,,
}
we can see that $\abs{A}=r$ acts as a dynamical order parameter for the mechanical resonators, in the sense of Kuramoto \cite{kuramoto:1975},
\eq{
 A(\ta) & = \f{N}\sum_iA_i(\ta) & \imp && \abs{A}\ei{\thet} & = \f{N}\sum_ir_i\ei{\thet_i} \,.
}
As before, we can use Bessel functions to work out the cavity amplitude response
%\eq{
%  \al=\ei{\ps(t)}\sum_mB_m\ei{m\om t} \,,
% }
% where
% \eq{
%  \ps(t) = -\f[2N\abs{A}]{\om}\cos\<{\om t+\thet} \hspace{5mm}\txt{and}\hspace{5mm} B_m = -\fr{\ii^{m+1}\ep J_m\<{\f[2N\abs{A}]{\om}}}{\bar{\ka}+\ii m\om} \,,
% }
and substitute this back into the equations for the individual oscillators to give the amplitude equations,
\eq{
 \D{A_i}{\ta} = -\<{1 + \De\om_i}A_i + N_iG\bar{\ep}^2\ei{\thet}\sum_{m=0,\infty}a_{m}\<{\bar{\de},\om}J_m\<{\f[2N\abs{A}]{\om}}J_{m+1}\<{\f[2N\abs{A}]{\om}} \,,
}
where $a_m\<{\bar{\de},\om}=a_{mr}\<{\bar{\de},\om} + \ii a_{mi}\<{\bar{\de},\om}$ is defined in the previous section.

\subsubsection{Two sets of nonidentical mechanical resonators}

In terms of two sets of oscillators this becomes
\eq{
 \D{A_1}{\ta} & = -\<{1 + \ii\De\om_1}A_1 + GN_1\bar{\ep}^2\<{A_1+A_2}F\<{\abs{A_1+A_2}} \,, \\
 \D{A_2}{\ta} & = -\<{1 + \ii\De\om_2}A_2 + GN_2\bar{\ep}^2\<{A_1+A_2}F\<{\abs{A_1+A_2}} \,.
}
If the $\De\om_i$ are equal they do not effect the radial motion and we still have
\eq{
 \D{r}{\ta} = - r +G\bar{\ep}^2 Nr F_r\<{Nr,\om,\de} \,,
}
which implies that  $N^2r^2=r_1^2+r_2^2+2r_1r_2\cos (\thet_2-\thet_1)$ is a constant of the motion. Substituting this into the equations for $A_i$ results in a linear system whose  symmetrical solution $N_1A_2=N_2A_1$ is stable. So apart from some transients the individual oscillators synchronize, $\Dt{\<{N_1A_2-N_2A_1}}=-\<{\ga+\ii\De\om}\<{N_1A_2-N_2A_1}$, as noted before. 

If the $\De\om_i$ are not equal the dynamics of the system, which is a function of the relative phase $\ph=\thet_2-\thet_1$ only, is given by the nonlinear system
\eq{
 \D{r_1}{\ta} & = -r_1 + \bar{\ep}^2GN_1\<{r_1F_r\<{Nr} + r_2\<{F_r\<{Nr}\cos\ph-F_i\<{Nr}\sin\ph}} \,, \\
 \D{r_2}{\ta} & = -r_2 + \bar{\ep}^2GN_2\<{r_2F_r\<{Nr} + r_1\<{F_r\<{Nr}\cos\ph+F_i\<{Nr}\sin\ph}} \,, \\
 \D{\ph}{\ta} & = \De\om_{21} + \bar{\ep}^2GF_i\<{Nr}\<{\<{N_2-N_1} + \<{\f[N_2r_1]{r_2}-\f[N_1r_2]{r_1}}\cos\ph} + \bar{\ep}^2GF_r\<{Nr}\<{\f[N_2r_1]{r_2}+\f[N_1 r_2]{r_1}}\sin\ph \,,
}
where $F_{i,r}\<{Nr}=F_{i,r}\<{Nr,\om,\de}$, $Nr=\abs{A_1+A_2}=\sqrt{r_1^2+r_2^2+2r_1r_2\cos\ph}$ and $\De\om_{21}=\De\om_{2}-\De\om_1$. For $N_1=N_2$ we can assume that $\De\om_{21}>0$ as the transformation  $(\De\om_{21} \to -\De\om_{21},\, \ph \to -\ph)$ and $(r_1 \to r_2$ and vice versa) leaves the equations unchanged. The coupling, however is strong rather than weak and the system  cannot be reduced to a phase model.  But it is nevertheless useful to compare our results with those of similar phase and phase amplitude models, \cite{pikovsky:2003,kuramoto:1975,lifshitz:2008,aronson:1990,strogatz:2000}. 

In the simplest two-oscillator phase model ($\dot{\ph}= \De\om -K \sin\ph$ with $\ph=\thet_2-\thet_1$) there are two critical points, approximately an ina-phase and an out-of-phase solution. One of the critical points is stable, for $\abs{\De\om}$ sufficiently small ($\abs{\De\om}<K$). Unsynchronized motion occurs when the critical points are lost via a saddle node bifurcation ($\abs{\De\om}>K$).  More complex models include a $\sin2\ph$ term in which case the in-phase solution ($\ph\approx0$) may loose stability to a stable  out-of-phase solution ($\ph\approx\pi$). The  model here can also be discussed in terms of  the stabilities of in-phase and out-of-phase solutions. However the 'unsynchronized behaviour' occurs as a transient state, resembling the transient rotational motion of a damped nonlinear pendulum started near to the separatrix of the undamped system. Similar motion has been noted for other systems with multi stability \cite{mueller:2008}. 

\begin{figure}[!htbp]\begin{center}
\includegraphics[scale=0.8]{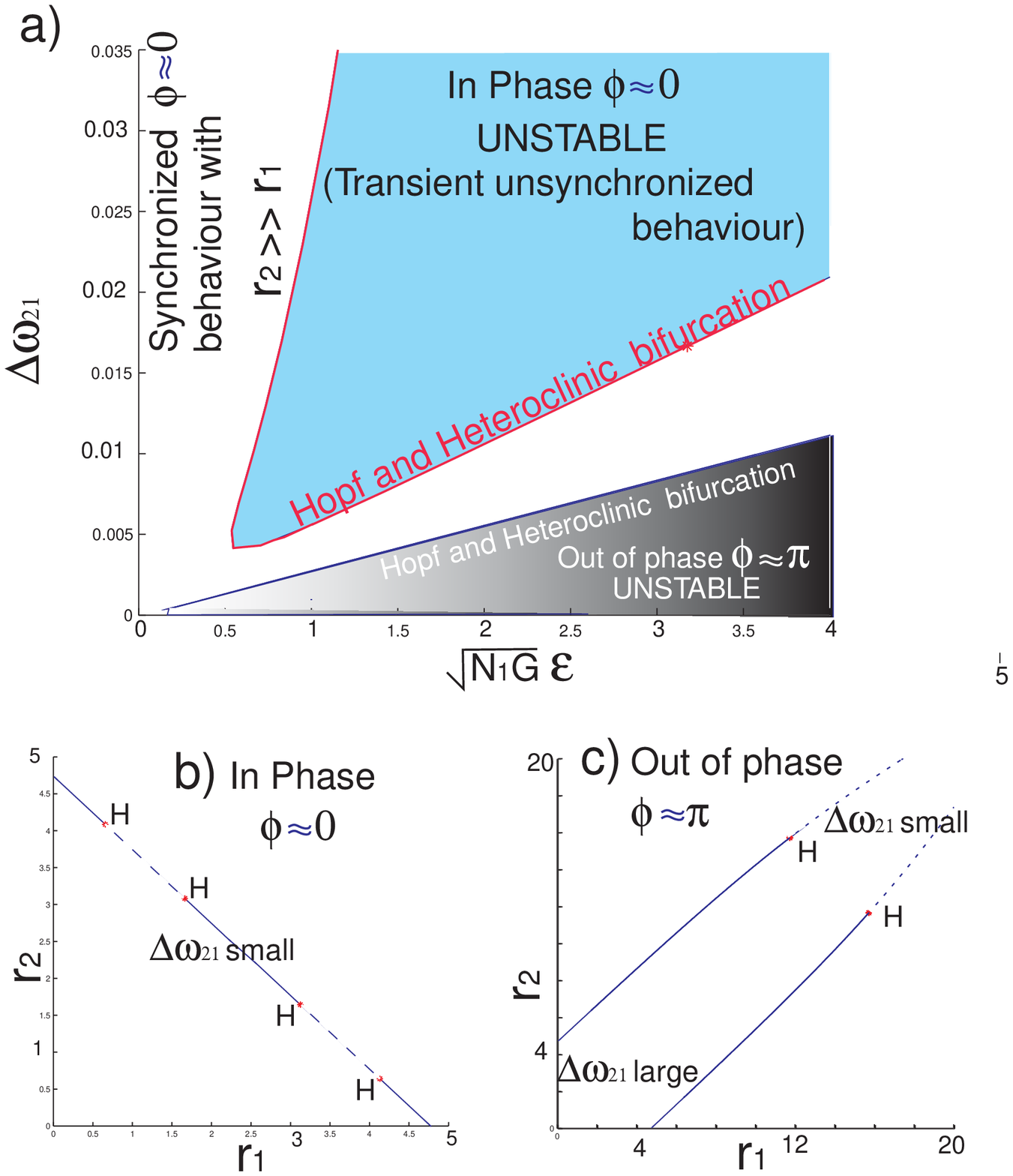}
\caption{The bifurcation diagram for 2 mechanical resonators for $N_1=N_2,\, \ga=0.0001,\,\om=2$ and $\de=-1.5$. The in-phase solutions are stable outside the shaded cyan regions.  The out-of-phase solutions are stable outside the slice near the horizontal axis. They are  singular at $\De\om_{21}=0$ and  unstable for $\abs{\De\om_{21}}$ small, where they occur for  very large values of $r_i$. In the unshaded regions both in-phase and out-of-phase solutions are stable, but have different basins of attraction. b)  shows the in-phase solution in $(r_1,\,r_2)$ space as $\De\om_{21}$ is varied. $r_1+r_2$ remains approximately constant. c) shows the out of phase solution in $(r_1,\,r_2)$ space as $\De\om_{21}$ is varied.}
\label{f:figure7}
\end{center}\end{figure}

Nonzero $\De\om_{21}$ breaks the symmetry and the in-phase critical points, which are still stable for $\De\om_{21}$ very small, only exist with $r_1\ne r_2$. Their relative sizes as $\abs{\De\om_{21}}$ is varied in shown in \fref{f:figure7} b). As $\abs{\De\om_{21}}$ is increased they loose stability via a Hopf bifurcation, \fref{f:figure7} a). This creates a stable periodic orbit which does not initially envelope the origin. However, in a bifurcation scenario typical of large amplitude coupling \cite{pikovsky:2003}, it grows rapidly to enclose the origin. (In ($r_1$,$r_2$,$\ph$) space this transition is a heteroclinic bifurcation with saddles at $r_{1\txt{ or }2}=0,\, \ph=\pm\f[\pi]{2}$.) Transient unsynchronized motion results for solutions started near the (unstable) in-phase solution, where solutions appear unbounded in-phase, but eventually become trapped by a stable out-of-phase solution. (In fact the out-of-phase solutions are only unstable for $\De\om_{21}$ very small, where they exist at large amplitude, \fref{f:figure7} c)). The bifurcation diagram \fref{f:figure7} was created using the package \MatCont with $F \<{\abs{A}^2,\bar{\ka},\om}$ approximated by the first four terms in its Maclaurin series in $\abs{A}^2$, $N_1=N_2$, $\ga=0.0001$, $\om=2$ and $\de=-1.5$.

\begin{figure}[!htbp]\begin{center}
\includegraphics[scale=1]{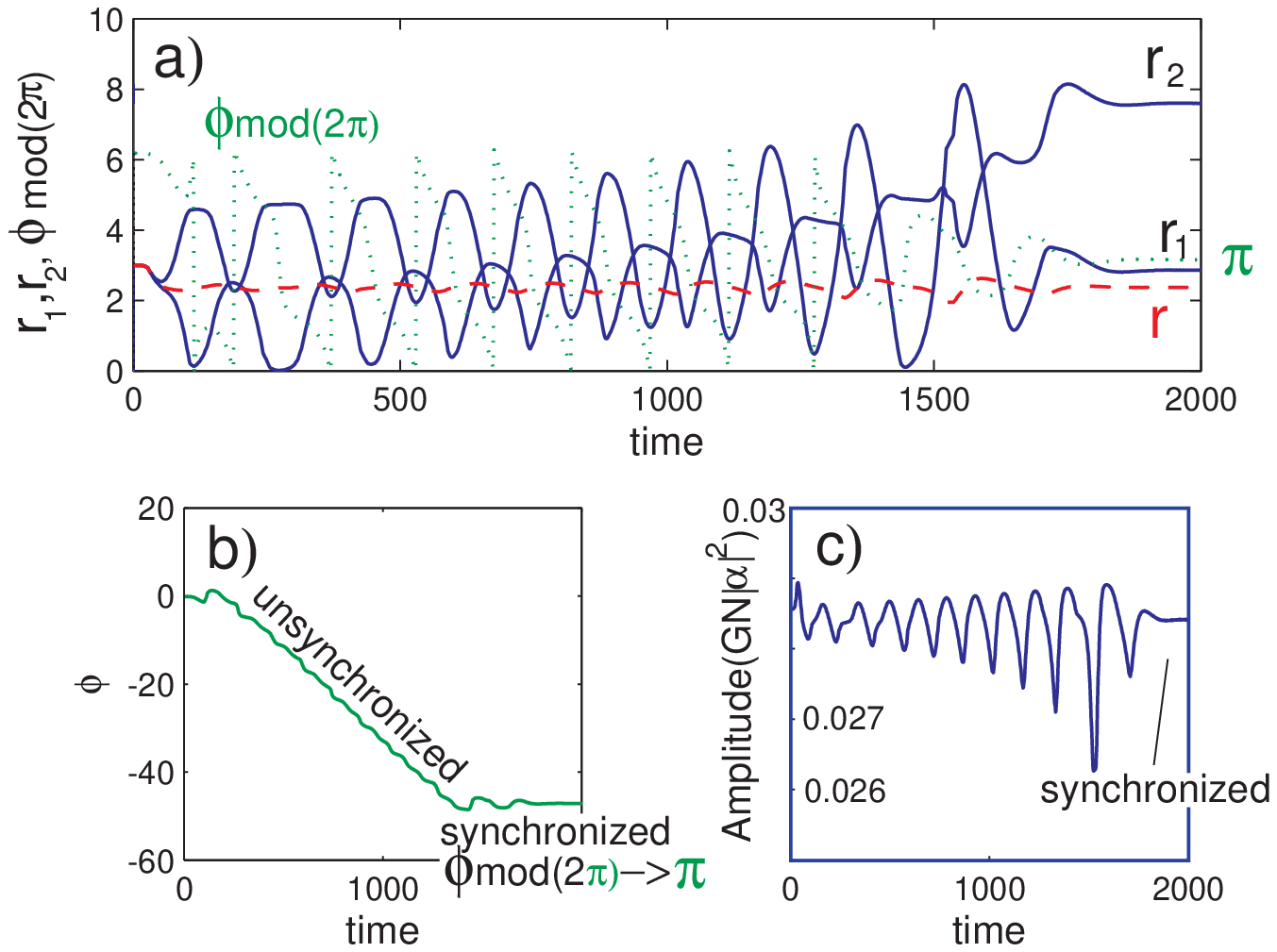}
\caption{Transient unsynchronized  motion for 2 nonidentical mechanical resonators for  $N_1=N_2,\,\om=2,\,\de=-1.5,\,\ga=0.0001,\,\sqrt{N_1G}\ep=2$ and $\De\om_{21}=0.04$. Started near the in-phase solution, in the blue shaded region of \fref{f:figure7}, the transient unsynchronized motion is only temporary. Eventually solutions are trapped by an out-of-phase solution ($\ph$ mod ($2\pi)\to \pi$). The variables are plotted against time. In a) $r_1$ and $r_2$ are shown in solid lines, the collective variable $r$ is dashed and $\ph$mod($2\pi)$ is dotted. b) is a plot of $\ph$ and c) is a plot of the amplitude of $GN\abs{\al}^2$.}
\label{f:figure8}
\end{center}\end{figure}

If we consider only the solutions started near the in-phase solution then, for sufficiently large $\sqrt{N_1G}\ep>0.5$, increasing $\abs{\De\om_{21}}$ engenders a loss of  synchronization, see \fref{f:figure8}. A heteroclinic bifurcation provides the real boundary for loss of synchronization and  eventually solutions synchronize into an out-of-phase solution. In the unsynchronized behaviour the radii  execute fairly large oscillations. However the oscillations in  the cavity amplitude  are  not large. A typical example is shown in \fref{f:figure8} for $\om=2,\,\de=-1.5,\,\ga=0.0001,\,\sqrt{N_1G}\ep=2$ and $\De\om_{21}=0.04$ starting near the unstable in-phase solution.

If the $N_i$ are not equal the bifurcation diagram is not symmetrical in $\De\om_{21}$. However apart from this it is not dissimilar. The in-phase  solution with $\ph=0 $ occurs for $\f[r_1]{r_2}=\f[N_1]{N_2}$ and looses stability as $\abs{\De\om_{21}}$ is increased away from zero, eventually stabilizing on an out of phase solution.

\subsubsection{Three sets of nonidentical mechanical resonators}

\begin{figure}[!htbp]\begin{center}
\includegraphics[scale=1]{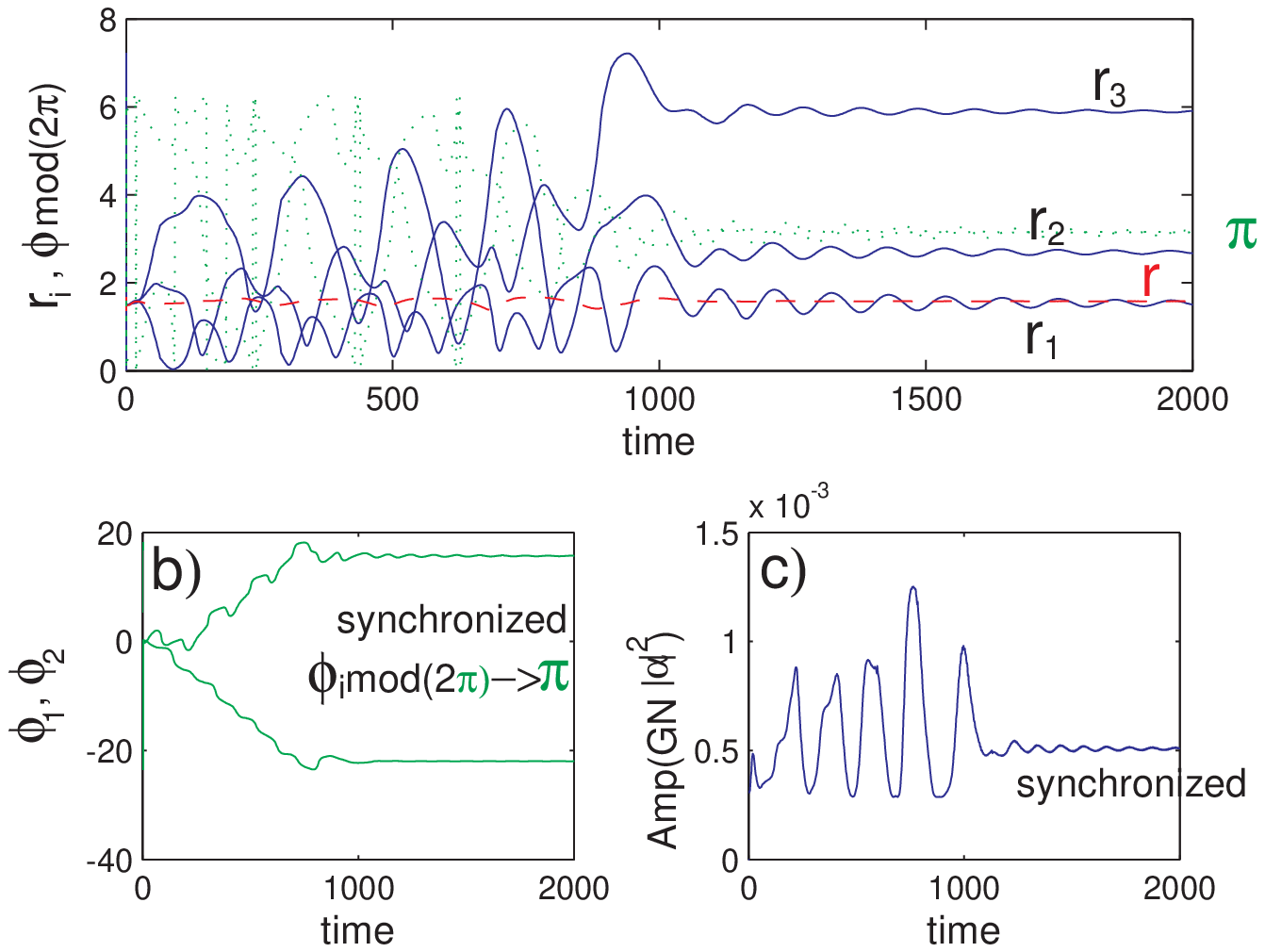}
\caption{ Transient unsynchronized motion of the in-phase solution for 3 sets of  mechanical nonidentical resonators for  $N_i$ equal, $\om=2,\,\de=-1.5,\,\ga=0.0001,\,\sqrt{N_iG}\ep=2$, $\De\om_{21}=0.04$ and $\De\om_{21}=0.045$.  Eventually solutions are trapped by an out-of-phase solution ($\ph_i$ mod ($2\pi)\to \pi$ here). The variables are plotted against time. In a) $r_i$  are shown in solid lines, the collective variable $r$ is dashed and $\ph_i\txt{ mod }\<{2\pi}$ are dotted. b) is a plot of $\ph_i$ and c) is a plot of the amplitude of $GN\abs{\al}^2$. }
\label{f:figure9}
\end{center}\end{figure}

The system for $N$ sets of mechanical resonators 
\eq{
 \D{r_i}{\ta}     = - r_i + \ep^2N_iG\pb{\sum_jr_j\<{F_r\<{Nr}\cos\<{\thet_j-\thet_i} - F_i\<{Nr}\sin\<{\thet_j-\thet_i}}} \,, \\
 \D{\thet_i}{\ta} = \De\om_{i} + \ep^2N_iG\pb{\sum_j\f[r_j]{r_i}\<{F_i\<{Nr}\cos\<{\thet_j-\thet_i} + F_r\<{Nr}\sin\<{\thet_j-\thet_i}}} \,,
}
where $\<{Nr}^2=\abs{\sum_{i=1}^NA_i}^2=\sum_{i,j=1}^Nr_ir_j\cos\<{\thet_i-\thet_j}$, may be reduced to  $2N-1$ equations of motion because the equations above are only functions of the relative phase: $\ph_i=\thet_{i+1}-\thet_i$. So three mechanical resonators  are described by five equations of motion for $r_1,\,r_2,\,r_3,\,\ph_1,\,\ph_2$. If the $\De\om_{i}$ are equal the model can be reduced to that for a single collective variable. In fact if any two of the $\De\om_{i}$ are equal then those two resonators can be thought of as one. Using the notation $\De\om_{ij}=\De\om_{i}-\De\om_{j}$, the three oscillator case reduces to the two oscillator case if  $\De\om_{21}=0$ or if $\De\om_{32}=0$ or if  $\De\om_{21}+\De\om_{32}=0$. \Fref{f:figure9} shows a typical example of loss of synchronization for  $\De\om_{21}+\De\om_{32}$ small. 

% The collective variable $r$, or functions of $r$, such as the amplitude of the oscillations in the cavity amplitude (amplitude of $GN\abs{\al}^2)$, are good indicators of synchronized behaviour. If the system is synchronized, either in an in-phase or an out-of-phase solution, $r$ is a constant. 
 
\begin{figure}[!htbp]\begin{center}
\includegraphics[scale=0.9]{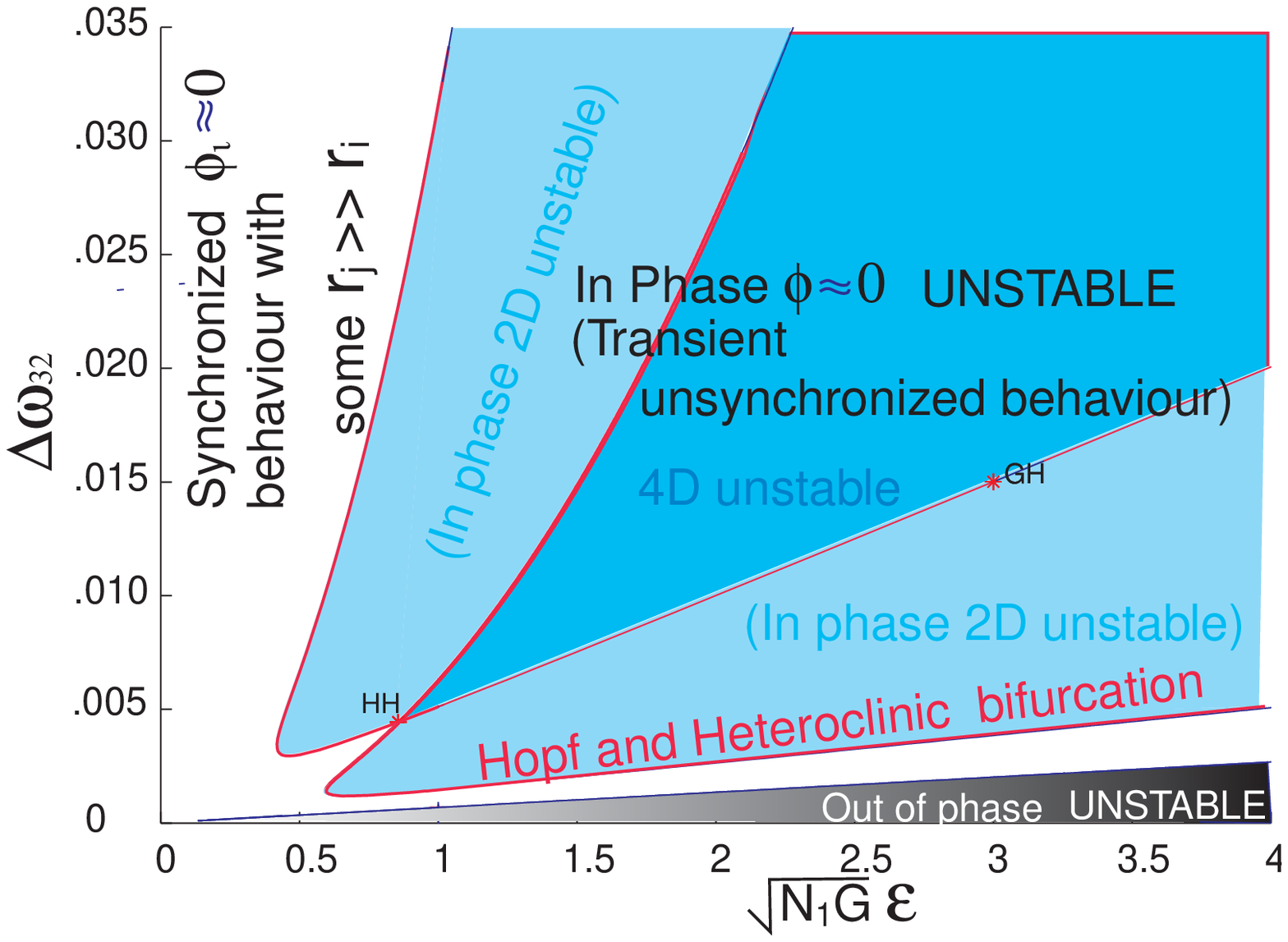}
\caption{The bifurcation diagram for three mechanical resonators for  $N_i$ equal, $\om=2,\,\de=-1.5,\,\ga=0.0001$, $\De\om_{21}=5\De\om_{32}$. The in-phase solution is stable outside the shaded cyan and blue  regions.  The out-of-phase solutions are stable outside the grey regions. In the blue region the in-phase solution is unstable to both $\ph_1$ and $\ph_2$ and the motion is transiently unsynchronized, eventually setting on an out-of-phase solution.  Once again the out-of-phase solutions tend to exist where at least some of the $r_i$ take larger values.}
\label{f:figure10}
\end{center}\end{figure}

From a dynamical point of view the three resonator case has only one in-phase motion ($\ph_i\approx0$) and one out-of-phase motion with $\ph_1\approx\pi,\,\ph_2\approx0$ or the other way round. (The case with both $\ph_i\approx\pi$ is dynamically the same as $\ph_1\approx \pi,\,\ph_2\approx 0$.) So as before we can think in terms of the in-phase and out-of-phase solutions. (This is not the case for $N \ge 4$.) Otherwise the bifurcation diagram is more complicated involving two sets of Hopf curves, however if $\De\om_{21}$ and $5\De\om_{32}$ are close the Hopf curves are also close. In contrast if they differ, as shown in \fref{f:figure10} where we have taken $\De\om_{21}=5\De\om_{32}$, three unstable regions result. The most complex behaviour occurs in the blue region in which  the in-phase solution is unstable to both $\ph_i$ and the motion may switch from librational to rotational motion in one or both of the $\ph_i$ apparently randomly. 
% Once again the bifurcations are of Hopf type with heteroclinic bifurcations very close by. Note that for larger values of $\sqrt{N_iG}\ep$ the upper bound for synchronized behaviour occurs approximately on a line in $(\sqrt{N_iG}\ep,\, \De\om_{32})$ space.

\section{Quantum mechanical description}\label{s:quantum}

\subsection{Quantum mechanical model}

The classical model derived in section \ref{s:MO_i} and analysed in section \ref{s:MO_sc} is given by the classical Hamiltonian
\eq{
 \ham = \hb\de\abs{\al}^2 + \sum_{i=1}^N\hb\om_i\abs{\al}^2 + \hb\<{\ep^*\al+\ep\al^*} + \sum_{i=1}^N\hb g_i\abs{\al}^2x_i \,,
}
where the microwave cavity amplitude $\al = x_c + \ii y_c$, and the dimensionless canonical positions $x_c$ and $x_i$, and their conjugate momenta $y_c$ and $y_i$ respectively, satisfy the Poisson bracket relations
\eq{
 \PB{x_c}{y_c} & = \f{2\hb} \,, \\
 \PB{x_i}{y_j} & = \f{2\hb}\de_{ij} \,.
}
The original canonical positions $\Ph=\sqrt{2E_cL}\;x_c$ and $q_i=\sqrt{\f[2E_m]{m\om^2}}\;x_i$, and their conjugate momenta $Q=\sqrt{2E_cC_0}\;y_c$ and $p_i=\sqrt{2mE_m}\;y_i$ respectively, satisfy the canonical Poisson bracket commutation relations
\eqa{
 \PB{\Ph}{Q}   & = 2\hb\PB{x_c}{y_c} && = 1 \,, \\
 \PB{q_i}{p_j} & = 2\hb\PB{x_i}{y_j} && = \de_{ij} \,.
}
The quantum mechanical description of the Hamiltonian dynamics matches that obtained by canonical quantisation of the classical Hamiltonian. We promote the canonical position and momenta $\iPh$ and $Q$ of the microwave cavity to the quantum mechanical operators $\qo{\iPh}$ and $\qo{Q}$ respectively. Similarly, we promote the nano-mechanical resonator positions $q_i$ and momenta $p_i$ to the quantum mechanical operators $\qop[i]$ and $\pop[i]$ respectively. We then define the annihilation operators for the microwave cavity field mode $\aop$, and the nano-mechanical vibrational modes $\bop[i]$. We again have the dimensionless microwave cavity quadrature operators $\xop[c]=\f{2}\<{\aop+\ad}$ and $\yop[c]=-\ii\f{2}\<{\aop-\ad}$, and the dimensionless nano-mechanical positions and momenta $\xop[i]=\f{2}\<{\bop[i]+\bd[i]}$ and $\yop[i]=-\ii\f{2}\<{\bop[i]-\bd[i]}$ respectively. We have the commutation relations for the quantum operators
\eq[e:MO_com]{
 \C{\aop}{\ad}       & = \Iop \,,         & \C{\xop[c]}{\yop[c]} & = \ii\f{2}\Iop \,,         & \C{\qo{\iPh}}{\qo{Q}} & = \ii\hb\Iop \,, \\
 \C{\bop[i]}{\bd[j]} & = \de_{ij}\Iop \,, & \C{\xop[i]}{\yop[j]} & = \ii\f{2}\de_{ij}\Iop \,, & \C{\qop[i]}{\pop[i]}  & = \ii\hb\de_{ij}\Iop \,, \\
}
in terms of which the corresponding quantum Hamiltonian in the interaction picture is given by
\eq[e:MO_iham]{
 \qham = \hb\de\<{\ada+\f{2}} + \sum_{i=1}^N\hb\om_i\<{\bdb[i]+\f{2}} + \hb\<{\ep^*\aop+\ep\ad} + \sum_{i=1}^N\hb g_i\<{\ada+\f{2}}\xop[i] \,.
}
For a realistic device we adopt a dissipative model. We model both the microwave cavity resonator and the mechanical resonators as being damped in zero temperature heat baths. A zero temperature heat bath for the cavity is certainly justified as the typical microwave cavity is at mK temperature and thus very close to zero \cite{wallraff:2004}. The $N$ zero temperature heat baths for the $N$ nano-mechanical resonators are not as good an approximation. However, the mean thermal occupation of the $i$th bath $\bar{n_i}\ne0$ does not enter the semi-classical equations, and thus the semi-classical bifurcation structure studied in \sref{s:MO_sc} is the correct one. The amplitude decay for the microwave cavity is $\ka$, and for the $i$th nano-mechanical resonator is $\ga_i$. We then describe the dissipative dynamics with the master equation (with weak damping and the rotating wave approximation for the system-environment couplings)
\eq[e:MO_me]{
  \Dt{\dm} = -\f[\ii]{\hb}\C{\qham}{\dm} + \ka\<{2\aop\dm\ad-\ada\dm-\dm\ada} + \sum_{i=1}^N\ga_i\<{2\bop[i]\dm\bd[i]-\bdb[i]\dm-\dm\bdb[i]} \,,
}
where $\dm$ is the density matrix of the coupled system.

Corresponding to the the classical description, we are interested in the $M$ collective quantities $\Xop[i]$ and $\Yop[i]$ defined by
\eq{
 \Xop[i] & = \f{N_i}\sum_{j\in S_i}g_j\xop[j] \,, \\
 \Yop[i] & = \f{N_i}\sum_{j\in S_i}g_j\yop[j] \,.
}
We can define creation and annihilation operators for these collective mechanical modes,
\eq{
 \Bop[i]    & = \f{N_i}\sum_{j\in S_i}g_j\bop[j] \,, \\
 \Bop[i]\ct & = \f{N_i}\sum_{j\in S_i}g_j\bd[j]  \,,
}
where from \eqref{e:MO_com}, we can show that the commutation relations for the new collective operators are
\eq{
 \C{\Bop[i]}{\Bd[j]} = \f[G_i]{N_i}\de_{i,j}\Iop \,.
}

\subsection{Fokker-Planck like equation}

From the Master equation \eqref{e:MO_me}, we proceed by deriving a Fokker-Planck like equation for the nano-electromechanical system which is the equation of motion of the Positive P function $P(\gma{\ch})$. The Positive P function is the Fourier Transform of the expectation of the normally-ordered characteristic function,
\eq{
  P(\gma{\ch}) = \f{\<{2\pi}^{2M+2}}\int\E{\ei{\la_{2M+2}\Bd[M]}\ei{\la_{2M+1}\Bop[M]}\cdots\ei{\la_4\Bd[1]}\ei{\la_3\Bop[1]}\ei{\la_2\ad}\ei{\la_1\aop}}\ee^{-i\gma{\la}\cdot\gma{\ch}}\dd{\gma{\la}} \,,
} % ### AMS -> IOP: flushed left
where
\eq{
 \gma{\ch} & = \bm{\al & \be & \mu_1 & \nu_1 & \mu_2 & \nu_2 & \cdots & \mu_M & \nu_M}^T \,, \\
 \gma{\la} & = \bm{\la_1 & \la_2 & \cdots & \la_{2M+2}}^T \,.
}
We follow the procedure outlined in \cite{walls:2008}. Using the appropriate commutation relations, we arrive at the Fokker-Planck like equation
\eq[e:MO_fokkerplanck]{
  \Dt{P(\gma{\ch})} = -\sum_i\pD{}{\ch_i}\pb{\ma{A}(\gma{\ch})}_iP(\gma{\ch}) + \f{2}\sum_{ij}\pD{}{\ch_i}\pD{}{\ch_j}\pb{\ma{B}(\gma{\ch})\ma{B}(\gma{\ch})^T}_{ij}P(\gma{\ch}) \,, \\
}
where the drift term vector $\ma{A}(\gma{\ch})$ is
\eq{
 \ma{A}(\gma{\ch})                    = \bm{-\ii\ep - \<{\ka+\ii\de}\al - \ii\f{2}\al\sum_{i=1}^MN_i\<{\mu_i+\nu_i} \\
                                             \ii\ep - \<{\ka-\ii\de}\be + \ii\f{2}\be\sum_{i=1}^MN_i\<{\mu_i+\nu_i} \\
                                            -\<{\ga+\ii\om_1}\mu_1 - \ii\f[G_1]{2}\al\be \\
                                            -\<{\ga-\ii\om_1}\nu_1 + \ii\f[G_1]{2}\al\be \\
                                            \vdots \\
                                            -\<{\ga+\ii\om_M}\mu_M - \ii\f[G_M]{2}\al\be \\
                                            -\<{\ga-\ii\om_M}\nu_M + \ii\f[G_M]{2}\al\be} \,,
}
and the diffusion term matrix $\ma{B}(\gma{\ch})\ma{B}(\gma{\ch})^T$ is
\eq{
 \ma{B}(\gma{\ch})\ma{B}(\gma{\ch})^T = \bm{0                   & 0                   & -\ii\f[G_1]{2}\al & 0                  
                                          & \cdots              & -\ii\f[G_M]{2}\al & 0 \\
                                            0                   & 0                   & 0                   & \ii\f[G_1]{2}\be 
                                          & \cdots              & 0                   & \ii\f[G_M]{2}\be \\
                                            -\ii\f[G_1]{2}\al & 0                   & 0                   & 0                  
                                          & \cdots              & 0                   & 0 \\
                                            0                   & \ii\f[G_1]{2}\be  & 0                   & 0                  
                                          & \cdots              & 0                   & 0 \\
                                            \vdots              & \vdots              & \vdots              & \vdots           
                                          & \ddots              & \vdots              & \vdots \\
                                            -\ii\f[G_M]{2}\al & 0                   & 0                   & 0                  
                                          & \cdots              & 0                   & 0 \\
                                            0                   & \ii\f[G_M]{2}\be  & 0                   & 0                  
                                          & \cdots              & 0                   & 0 } \,.
}

If we consider only the drift term of the Fokker-Planck like equation \eqref{e:MO_fokkerplanck}, and make the mappings $\be\mapsto\al^*$ and $u\mapsto v^*$ to reduce the phase space dimension by half onto the semi-classical phase space (the positive P function has twice the dimensionality of the classical phase space), then we obtain the semi-classical equations of motion.
%\eq[e:MO_sceom]{
% \Dt{\al} & = -\ii\de\al - \ii\ep - \ii\al\sum_{i=1}^MN_iX_i - \ka\al \,, \\
% \Dt{X_i} & = \om_iY_i - \ga_iX_i \,, \\
% \Dt{Y_i} & = -\om_iX_i - \f[G_i]{2}\abs{\al}^2 - \ga_iY_i \,,
%}
%where $i=1,\ldots,M$. These equations match the equations of motion we would generate by factoring moments $\E{\aop\Xop}=\E{\aop}\E{\Xop}$ in the quantum equations of motion of operator expectations, \eqref{e:MO_qeom}. With this factorisation, the mappings from expectation values of quantum operators to semi-classical dynamic variables is then $\E{\aop}\mapsto\al$, $\E{\Xop[i]}\mapsto X_i$, $\E{\Yop[i]}\mapsto Y_i$, and $\E{\Xop}\mapsto X=\sum_{i=1}^M\f[N_i]{N}X_i$. The next section is devoted to the behaviour of the semi-classical nano-electromechanical system.

\subsection{Quantum spectra}\label{s:MO_spectra}

A future direction for research, that builds on the work of this paper, is the investigation of the quantum physics associated with the multi-stable semi-classical limit cycles. As a starting point, in this section we calculate the linearised spectrum as we increase the driving strength to approach the first Hopf bifurcation at the supercritical Hopf line for blue detuning ($\de<0$) in \Fref{f:figure2} and \Fref{f:figure3}. We do this calculation for the case of a single group of nano-mechanical resonators following the procedure of \cite{holmes:2009}. For a single group, using the dimensionless notation where we have re-scaled the coupling coefficients and time by the cavity dissipation rate $\ka$, we have the stochastic differential equations of motion corresponding to the Fokker-Planck like equation \eqref{e:MO_fokkerplanck}
\eq{
 \Dt{\gma{\ch}} = \ma{A}(\gma{\ch}) + \ma{B}(\gma{\ch})\ma{E}(t) \,,
}
where
\eq{
 \gma{\ch} & = \bm{\al & \be & \mu & \nu}^T \,,
}
the drift term vector $\ma{A}(\gma{\ch})$ is
\eq{
 \ma{A}(\gma{\ch})                    = \bm{-\<{1+\ii\de}\al - \ii\f{2}\al N\<{\mu+\nu} - \ii\ep  \\
                                            -\<{1-\ii\de}\be + \ii\f{2}\be N\<{\mu+\nu} + \ii\ep \\
                                            -\<{\ga+\ii\om}\mu - \ii\f[G]{2}\al\be \\
                                            -\<{\ga-\ii\om}\nu + \ii\f[G]{2}\al\be} \,,
}
the diffusion term matrix $\ma{B}(\gma{\ch})\ma{B}(\gma{\ch})^T$ is
\eq{
 \ma{B}(\gma{\ch})\ma{B}(\gma{\ch})^T = \bm{0               & 0               & -\ii\f[G]{2}\al & 0 \\
                                            0               & 0               & 0               & \ii\f[G]{2}\be \\
                                            -\ii\f[G]{2}\al & 0               & 0               & 0 \\
                                            0               & \ii\f[G]{2}\be  & 0               & 0} \,,
}
and $\ma{E}(t)$ is the noise process. The principal matrix square root of the diffusion matrix $\ma{B}(\gma{\ch})\ma{B}(\gma{\ch})^T$ is
\eq{
 \ma{B}(\gma{\ch}) = \ma{B}(\gma{\ch})\mt = \f[\sqrt{G}]{2}\bm{\sqrt{\al}     & 0             & -\ii\sqrt{\al} & 0 \\
                                                               0              & \sqrt{\be}    & 0              & \ii\sqrt{\be} \\
                                                               -\ii\sqrt{\al} & 0             & \sqrt{\al}     & 0 \\
                                                               0              & \ii\sqrt{\be} & 0              & \sqrt{\be}} \,.
}
The diffusion matrix and its square root have determinants
\eq{
 \det\pB{\ma{B}(\gma{\ch})\ma{B}(\gma{\ch})\mt} & = \f{16}G^4\al^2\be^2 \,, \\
 \det\pB{\ma{B}(\gma{\ch})}                     & = \f{4}G^2\al\be \,,
}
and the two matrices are thus positive definite on the semi-classical manifold where $\be=\al^*$. We see that the off-diagonal terms with the factors of $\ii$ in the matrix square root $\ma{B}(\gma{\ch})$ will take the solution off the semi-classical manifold, and will lead to quantum correlations.

We will linearise these equations of motion about the semi-classical fixed points we obtained in \sref{s:MO_sc}. In terms of the stochastic differential equations above, we make the mappings $\be\mapsto\al^*$ and $u\mapsto v^*$ to reduce the phase space dimension by half onto the semi-classical phase space (the positive P function has twice the dimensionality of the classical phase space). The linearised stochastic differential equations are then
\eq{
 \Dt{\gma{\ch}} \approx \ma{M}\<{\gma{\ch}-\gma{\ch}_0} + \ma{D}^{\f{2}}\ma{E}(t) \,,
}
where our Jacobian matrix $\ma{M}$ is
\eq{
 \ma{M} = \pD{\ma{f}(\gma{\ch}_0)}{\gma{\ch}} = \bm{-\<{1+\ii\de} - \ii\f{2}N\<{\mu_0+\mu_0^*} & 0 & -\ii\f{2}\al_0  & -\ii\f{2}\al_0 \\
                                                    0 & -\<{1-\ii\de} + \ii\f{2}N\<{\mu_0+\mu_0^*} & \ii\f{2}\al_0^* & \ii\f{2}\al_0^* \\
                                                    -\ii\f[G]{2}\al_0^* & -\ii\f[G]{2}\al_0        & -\<{\ga+\ii\om} & 0 \\
                                                    \ii\f[G]{2}\al_0^*  & \ii\f[G]{2}\al_0         & 0               & -\<{\ga-\ii\om}} \,,
}
$X_0=\frac12(\mu_0+\mu_0^*)$ and our diffusion matrix about the semi-classical fixed points $\ma{D}=\ma{B}(\gma{\ch}_0)\ma{B}(\gma{\ch}_0)^T$ is
\eq[e:diffmatrix]{
 \ma{D} = \bm{0                 & 0                  & -\ii\f[G]{2}\al_0 & 0 \\
              0                 & 0                  & 0                 & \ii\f[G]{2}\al_0^* \\
              -\ii\f[G]{2}\al_0 & 0                  & 0                 & 0 \\
              0                 & \ii\f[G]{2}\al_0^* & 0                 & 0} \,.
}
The linearised normally ordered moments at steady state can be expressed in terms of these matrices \cite{walls:2008}
\eq{
 \ma{S}(\iOm) = \f{2\pi}\int_{-\infty}^\infty\emi{\iOm\ta}\E{\gma{\ch}(t)\gma{\ch}(t+\ta)^T}_{t\to\infty}\dd\ta = \f{2\pi}\<{\ii\iOm\ma{I}-\ma{M}}^{-1}\ma{D}\<{-\ii\iOm\ma{I}-\ma{M}^T}^{-1} \,.
}
We plot the microwave cavity component of these quantum noise spectra in \Fref{f:MO_spectrum}. We see that in \Fref{f:MO_spectrum} (a) as the Hopf bifurcation is approached, the spectrum becomes more sharply peaked at two frequencies. The frequency corresponding to the Hopf bifurcation --- the magnitude of the two purely imaginary eigenvalues --- is the peak at the mechanical frequency $\om$. The second, shorter but broader peak, is at the detuning $\de$. For a drive detuned exactly on a sideband, these two peaks coincide. Beyond the supercritical Hopf bifurcation, the semi-classical fixed point is no longer stable and we enter the regime dominated by the first stable limit cycle, where we have the oscillatory motion analysed by semi-classical amplitude equations in \sref{s:MO_sc}. However, we can continue to linearise about this point, and show the results in \Fref{f:MO_spectrum} (b). The two peaks begin to converge as the driving strength and coupling are increased.

\fig[0.7]{f:MO_spectrum}{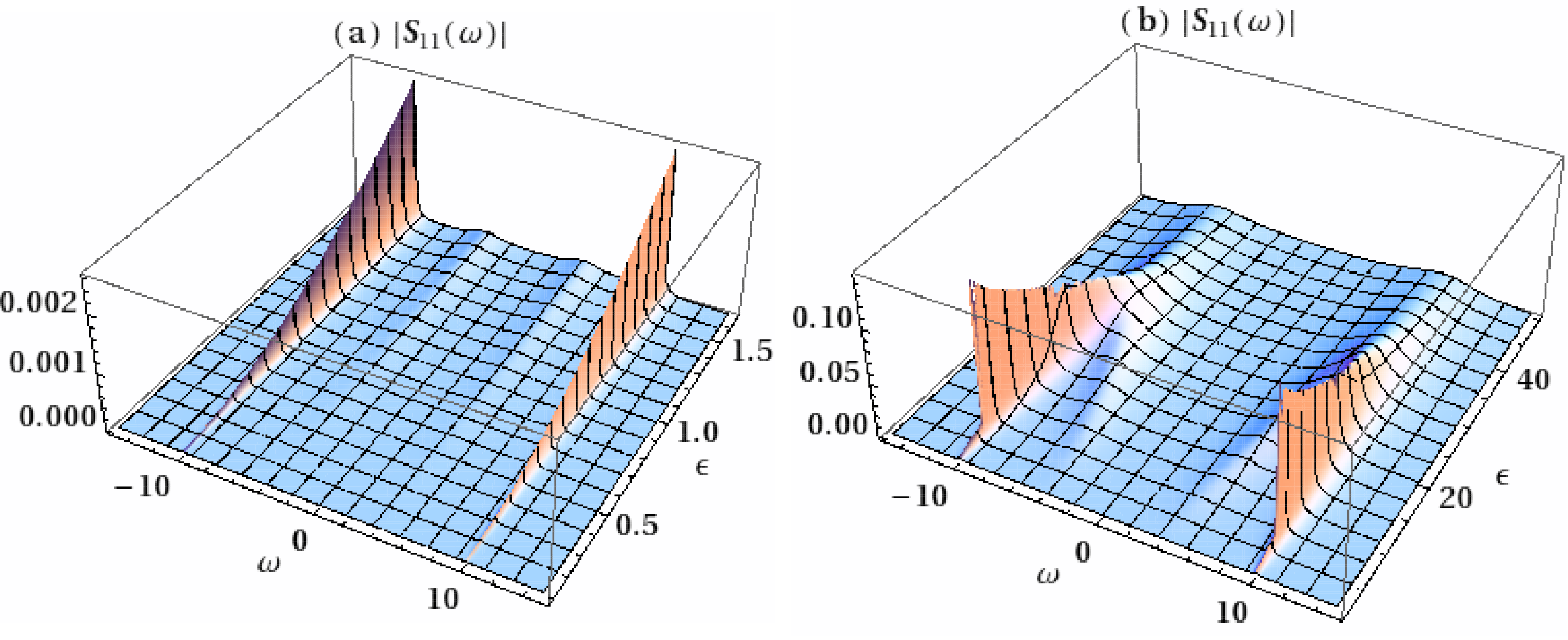}{
 Linearised quantum noise spectrum of the microwave cavity $S_{11}(\iOm)$: (a) approaching the Hopf bifurcation; and (b) continuing the linearisation beyond the Hopf bifurcation. The magnitude of the normally ordered cavity spectrum at steady state $\f{2\pi}\int_{-\infty}^\infty\emi{\iOm\ta}\E{\al(t)\al(t+\ta)^T}_{t\to\infty}\dd\ta$, the first diagonal element of $\ma{S}(\iOm)$, is plotted at the frequency $\iOm$ for varying driving amplitude $\ep$. Here we have set $\om=10$, $\de=-4$, $\ga=0.001$, $N=1$, $G=1$, for which the Hopf bifurcation occurs at a driving strength of $\ep_h\approx 1.76$. %from \eqref{e:MO_ehblue}.
}{Linearised quantum noise spectra of the microwave cavity approaching Hopf bifurcation}

The spectra calculated here correspond to the stationary fluctuations in the cavity field. The power spectrum of these fluctuations can be directly measured by homodyne detection.  Below the Hopf bifurcation, the noise power spectrum of these fluctuations is peaked at a frequency associated with the decay of fluctuations back to the fixed point. The width of the peaks gives the time scale of this decay. Above the Hopf bifurcation, the fluctuations decay onto the limit cycles. In our model, there are no thermal fluctuations and all fluctuations are due to intrinsic quantum noise manifest as off-diagonal components in the diffusion matrix in \eqref{e:diffmatrix}. A more careful study is required to determine if the multiple peak structure evident in \Fref{f:MO_spectrum}(b) is evidence for dissipative quantum switching between limit cycles.

\section{Discussion and Conclusion}\label{s:MO_c}
We have discussed the situation in which multiple mechanical resonators are coupled to a single  mode of the electromagnetic field in a microwave superconducting cavity.  This interaction results in an all-to-all coupling between each of the mechanical resonators that is highly nonlinear. However if the oscillators are identical, they synchronize and a collective variable can be used to understand the dynamics. Analysis of the dynamics of this collective variable (see the bifurcation diagram in \fref{f:figure2}) reveals the prevalence of periodic behaviour and suggests the use of amplitude equations \eqref{e:ampeqs} to describe the dominant oscillation.  Even though the amplitude equations involve elliptic functions their overall form is relatively simple for small mechanical damping and from them we are able to gain considerable insight into the dynamics of the collective variable. 

The form of the amplitude equations imply the presence of multiple periodic orbits and hysteresis (at the bifurcations of the periodic orbits). Specific results are also easy to extract; for instance we are able to plot the amplitudes of the mechanical resonators as a function of the external forcing for specific values of the other parameters (\fref{f:figure4}) and to locate the saddle node bifurcations of periodic orbits where hysteresis would occur as a result of a slight change in the mechanical forcing (\fref{f:figure2}, \fref{f:figure3}, and \fref{f:figure5}). 

The simplicity of the amplitude equations means that it is straightforward to extend the identical mechanical resonator case to one with distinct subgroups of identical oscillators. Considering two and three frequency subgroups we are able to give bifurcation diagrams showing the regions where synchronization occurs. Synchronization is lost via a mechanism involving a Hopf and heteroclinic bifurcation similar to that found in large amplitude forcing, rather than the sniper bifurcation that is involved in small amplitude forcing and, although in a reduced form, in Kuramoto's phase model. In spite of this difference there is a single collective variable $Nr$ that functions as a measure of synchronized behaviour and that is related to a measurable quantity, the cavity amplitude.

Given the current interest in fabricating nanomechanical resonators in microwave cavities, our model offers a realizable and very controllable way to study synchronization in a system with all-to-all coupling via a common field mode. While the equations for our model cannot be reduced to a simple phase model, it offers some advantages over more complex naturally occurring examples of synchronization.  A particularly important feature is that the measured quantity --- the cavity field leaving the microwave resonator ---  has an amplitude that is directly proportional to a collective parameter similar to the order parameter introduced in previous studies of synchronization. The need to use very low temperatures required for superconducting circuits may seem a disadvantage but in fact leads to a huge reduction in noise both for the mechanics and the microwave field. This should lead to especially clean observation of multi stability and perhaps even controlled switching between limit cycles. In the long run it also motivates us to study the effect of quantum noise on synchronization, and to look for quantum signatures of synchronization which will be the subject of a future paper.

\appendix %\section*{Appendix}\setcounter{section}{1}

\section{Experimental parameters}\label{s:MO_aexp}

A number of experiments are described by the model examined in this paper,
\eq{
 \Dt{\dm} = -\f[\ii]{\hb}\C{\qham}{\dm} + \ka\<{2\aop\dm\ad-\ada\dm-\dm\ada} + \sum_{i=1}^N\ga_i\<{2\bop[i]\dm\bd[i]-\bdb[i]\dm-\dm\bdb[i]} \,,
}
where
\eq{
 \qham = \hb\de\<{\ada+\f{2}} + \sum_{i=1}^N\hb\om_i\<{\bdb[i]+\f{2}} + \hb\<{\ep^*\aop+\ep\ad} + \sum_{i=1}^N\f{2}\hb g_i\<{\ada+\f{2}}\<{\bop[i]+\bd[i]} \,,
}
and
\eq{
 \C{\aop}{\ad}       & =        \Iop \,, \\
 \C{\bop[i]}{\bd[j]} & =        \de_{i,j}\Iop \,.
}  % ### AMS -> IOP: unaligned and made multi-line
A summary of the different values of the parameters for a selection of these experiments is given in \tref{t:MO_exp1}. In terms of the dimensionless parameters introduced in \sref{s:MO_sc} these become those listed in \tref{t:MO_exp2}. Note that the detuning $\de$ is typically set to be on a mechanical frequency sideband, such that $\de=\om_i$; and thus while not an experimental limitation, the range of $\de$ we list in the table is $\de\le\om_i$. Also note that the maximum driving $\abs{\ep}$ indicates the maximum driving before the cavity becomes nonlinear causing our model to fail. Finally, also note that the factors of $2$ in front of $\ka$ and $\ga_i$ in \tref{t:MO_exp1} are present because our $\ka$ and $\ga_i$ (as defined by the master equation above) are amplitude decay rates not occupation number decay rates.

\begin{table}[!h]\begin{center}\begin{tabular}{{@{}cccccccc@{}}}
 \toprule
 \multicolumn{2}{c}{Experiment} & \multicolumn{3}{c}{Mode $\aop$} & \multicolumn{2}{c}{Mode $\bop[i]$} & Coupling \\
 Ref. & Type & $\f[\om_c]{2\pi}\pb{\un{Hz}}$ & $2\f[\ka]{2\pi}\pb{\un{Hz}}$ & $\f[\abs{\ep}]{2\pi}\pb{\un{Hz}}$ & $\f[\om_i]{2\pi}\pb{\un{Hz}}$ & $2\f[\ga_i]{2\pi}\pb{\un{Hz}}$ & $\f[g_i]{2\pi}\pb{\un{Hz}}$ \\
 \midrule
 \cite{teufel:2009}      & S & $7.49\times10^9$   & $<2.88\times10^6$ &                           & $1.04\times10^6$  & $0.67$    & $866.7\times10^{-3}$ \\
 \cite{teufel:2008}      & S & $5.22\times10^9$   & $230\times10^3$   & $\lesssim2.145\times10^9$ & $1.53\times10^6$  & $<5.08$   & $190.7\times10^{-3}$ \\
 \cite{teufel:2008b}     & S & $4\times10^9$      & $400\times10^3$   &                           & $0.1\times10^6$   & $<1$      &\\
                         &   & $\to10\times10^9$  & $\to 1\times10^6$ &                           & $\to6\times10^6$  & $\to<6$   &\\
 \cite{teufel:2008b}     & S & $7.55\times10^9$   & $302\times10^3$   &                           & $1.41\times10^6$  & $<371.1$  &\\
 \cite{regal:2008}       & S & $\sim5\times10^9$  & $490\times10^3$   &                           & $2.3\times10^6$   & $19.2$    & $49.55\times10^{-3}$ \\
 \cite{sulkko:2010}      & S & $7.64\times10^9$   & $382\times10^3$   & $\lesssim2.434\times10^9$ & $67\times10^6$    & $248.1$   & $25.03$ \\
 \cite{thompson:2008}    & M & $282\times10^{12}$ & $4.07\times10^6$  &                           & $134\times10^3$   & $0.122$   & $27.8$  \\
 \cite{anetsberger:2009} & T &                    & $>4.9\times10^6$  &                           & $6.5\times10^6$   & $65$      & \\
                         &   &                    &                   &                           & $\to16\times10^6$ & $\to1600$ & \\
 \cite{anetsberger:2009} & T &                    & $50\times10^6$    &                           & $10.74\times10^6$ & $202.64$  & $147.3$ \\
 \cite{anetsberger:2009} & T &                    & $50\times10^6$    &                           & $8\times10^6$     & $200$     & $55.6$  \\
% \cite{chan:2009}        & C & $200\times10^{12}$ & $2.86\times10^7$  &                           & $1.7\times10^8$   &           &   \\
 \bottomrule
\end{tabular}\caption[Raw experimental coupling values for various systems]{Raw experimental coupling values for various systems. The `Type' column indicates the experimental context: `S' indicates a superconducting microwave coplanar waveguide resonator ($\aop$) coupled to a nano-mechanical resonator ($\bop[i]$); `M' indicates an optical cavity ($\aop$) coupled to a micro-mechanical membrane ($\bop[i]$); `T' indicates a toroidal micro-resonator ($\aop$) coupled to a nano-mechanical string resonator ($\bop[i]$); and `C' indicates an opto-mechanical crystal array where an optical mode of a cell ($\aop$) is coupled to a mechanical mode of a cell ($\bop[i]$).}\label{t:MO_exp1}\end{center}\end{table}

\begin{table}[!h]\begin{center}\begin{tabular}{{@{}ccccccc@{}}}
 \toprule
 \multicolumn{2}{c}{Experiment} & \multicolumn{2}{c}{Mode $\aop$} & \multicolumn{2}{c}{Mode $\bop[i]$} & \multicolumn{1}{c}{Coupling} \\
 Ref. & Type & $\abs{\de\pr}=\f[\abs{\de}]{\ka}$ & $\abs{\ep\pr}=\f[\abs{\ep}]{\ka}$ & $\om_i\pr=\f[\om_i]{\ka}$ & $\ga_i\pr=\f[\ga_i]{\ka}$ & $g_i\pr=\f[g_i]{\ka}$ \\
 \midrule
 \cite{teufel:2009}      & S & $\lesssim0.722$ &                          & $0.722$    & $2.33\times10^{-7}$     & $6.02\times10^{-7}$ \\
 \cite{teufel:2008}      & S & $\lesssim13.26$ & $\lesssim1.87\times10^4$ & $13.26$    & $2.209\times10^{-5}$    & $1.66\times10^{-6}$ \\
 \cite{teufel:2008b}     & S & $\lesssim0.5$   &                          & $0.5$      & $2.5\times10^{-6}$      & \\
                         &   & $\to\lesssim12$ &                          & $\to12$    & $\to6\times10^{-6}$     & \\
 \cite{teufel:2008b}     & S & $\lesssim9.34$  &                          & $9.34$     & $1.23\times10^{-3}$     & \\
 \cite{regal:2008}       & S & $\lesssim9.39$  &                          & $9.39$     & $3.92\times10^{-5}$     & $2.02\times10^{-7}$ \\
 \cite{sulkko:2010}      & S & $\lesssim350.8$ & $\lesssim1.28\times10^4$ & $350.8$    & $6.5\times10^{-4}$      & $1.31\times10^{-4}$ \\
 \cite{thompson:2008}    & M & $\lesssim0.066$ &                          & $0.066$    & $3\times10^{-8}$        & $1.37\times10^{-5}$ \\
 \cite{anetsberger:2009} & T & $<2.65$         &                          & $<2.65$    & $<1.33\times10^{-5}$    & \\
                         &   & $\to<6.53$      &                          & $\to<6.53$ & $\to<3.27\times10^{-4}$ & \\
 \cite{anetsberger:2009} & T & $\lesssim0.43$  &                          & $0.43$     & $4.05\times10^{-6}$     & $5.89\times10^{-6}$ \\
 \cite{anetsberger:2009} & T & $\lesssim0.32$  &                          & $0.32$     & $4\times10^{-6}$        & $2.22\times10^{-6}$ \\ 
% \cite{chan:2009}        & C & $\lesssim11.9$  &                          & $11.9$     &                         & \\
 \bottomrule
\end{tabular}\caption[Dimensionless experimental coupling values for various systems]{Dimensionless experimental coupling values for various systems. The `Type' column indicates the experimental context: `S' indicates a superconducting microwave coplanar waveguide resonator ($\aop$) coupled to a nano-mechanical resonator ($\bop[i]$); `M' indicates an optical cavity ($\aop$) coupled to a micro-mechanical membrane ($\bop[i]$); `T' indicates a toroidal micro-resonator ($\aop$) coupled to a nano-mechanical string resonator ($\bop[i]$); and `C' indicates an opto-mechanical crystal array where an optical mode of a cell ($\aop$) is coupled to a mechanical mode of a cell ($\bop[i]$).}\label{t:MO_exp2}\end{center}\end{table}

\acknowledgements
This work has been supported by the Australian Research Council.

\clearpage

\section*{References}

\bibliographystyle{apsrev}
\bibliography{meaney_refs_31aug2011}{}

\end{document}